\documentclass[twocolumn,prapplied,aps,superscriptaddress]{revtex4-1}
\usepackage[latin9]{inputenc}
\usepackage{mathtools}
\usepackage{amsbsy}
\usepackage{amstext}
\usepackage[unicode=true,pdfusetitle,
 bookmarks=false,
 breaklinks=false,pdfborder={0 0 1},backref=false,colorlinks=false]
 {hyperref}
\usepackage[dvipsnames]{xcolor}
 \hypersetup{colorlinks=true,citecolor=NavyBlue,filecolor=black,linkcolor=black,urlcolor=black}

\makeatletter

\@ifundefined{textcolor}{}
{%
 \definecolor{BLACK}{gray}{0}
 \definecolor{WHITE}{gray}{1}
 \definecolor{RED}{rgb}{1,0,0}
 \definecolor{GREEN}{rgb}{0,1,0}
 \definecolor{BLUE}{rgb}{0,0,1}
 \definecolor{CYAN}{cmyk}{1,0,0,0}
 \definecolor{MAGENTA}{cmyk}{0,1,0,0}
 \definecolor{YELLOW}{cmyk}{0,0,1,0}
}

\newcommand\ket[1]{\left|#1\right\rangle}

\usepackage{amsmath}
\usepackage{graphicx}
\usepackage{amssymb}
\usepackage{txfonts,color}
\usepackage{dsfont}
\makeatother

\begin{document}
\title{Versatile Atomic Magnetometry Assisted by Bayesian Inference}
\author{R. Puebla}
\affiliation{Instituto de F{\'i}sica Fundamental, IFF-CSIC, Calle Serrano 113b, 28006 Madrid, Spain}
\affiliation{Centre for Theoretical Atomic, Molecular and Optical Physics, Queen's University Belfast, Belfast BT7 1NN, United Kingdom}
\author{Y. Ban}
\affiliation{Department of Physical Chemistry, University of the Basque Country UPV/EHU, Apartado 644, 48080 Bilbao, Spain}
\affiliation{School of Materials Science and Engineering, Shanghai University, 200444 Shanghai, China}
\author{J. F. Haase}
\affiliation{Institute for Quantum Computing, University of Waterloo, Waterloo, Ontario, Canada, N2L 3G1}
\affiliation{Department of Physics \& Astronomy, University of Waterloo, Waterloo, Ontario, Canada, N2L 3G1}
\author{M. B. Plenio}
\affiliation{Institute of Theoretical Physics and IQST, Albert-Einstein-Allee 11, Universit\"at Ulm, D-89069 Ulm, Germany}
\author{M. Paternostro}
\affiliation{Centre for Theoretical Atomic, Molecular and Optical Physics, Queen's University Belfast, Belfast BT7 1NN, United Kingdom}
\author{J. Casanova}
\affiliation{Department of Physical Chemistry, University of the Basque Country UPV/EHU, Apartado 644, 48080 Bilbao, Spain}
\affiliation{IKERBASQUE, Basque  Foundation  for  Science,  Maria  Diaz  de  Haro  3,  48013  Bilbao, Spain}

\begin{abstract}
 Quantum sensors typically translate external fields into a periodic response whose frequency is then determined by analyses performed in Fourier space. This allows for a linear inference of the parameters that characterize external signals. In practice, however, quantum sensors are able to detect  fields only in a narrow range of amplitudes and frequencies.  A departure from this range, as well as the presence of significant noise sources and short detection times, lead to a loss of the linear relationship between the response of the sensor and the target field, thus limiting the working regime of the sensor. Here we address these challenges by means of a Bayesian inference approach that is tolerant to strong deviations from desired periodic responses of the sensor and is able to provide reliable estimates even with a very limited number of measurements. We demonstrate our method for an $^{171}$Yb$^{+}$ trapped-ion quantum sensor but stress the general applicability of this approach to different systems.
\end{abstract}

\maketitle

\section{Introduction}
Achieving efficient magnetometry is of considerable importance in a broad range of areas of fundamental and applied science~\cite{Lenz06,Edelstein07}. Nuclear magnetic resonance (NMR) techniques~\cite{Levitt08, Ernst87}, which led to important applications such as NMR spectroscopy~\cite{Gunter13}, magnetic resonance imaging~\cite{Plewes12}, and their recent
extensions to the nanoscale \cite{MullerKC+14,SchmittGS+17,SchwartzRS+19}, are specific examples that depend crucially on accurate and efficient magnetometry techniques. Other remarkable applications include magnetic force microscopy~\cite{Kazakova19}, which allows the scanning of thin materials for -- to throw an example -- magnetic recording~\cite{Bai04} and may achieve a spatial resolution of the order of tens of nanometers.  A new generation of devices that exploit quantum properties to characterize weak electromagnetic signals are superconducting quantum interference devices SQUIDs~\cite{Jaklevic64}. These possess excellent magnetic sensitivity and have dimensions ranging from microns~\cite{Cleuziou06} to tens of nanometers in the case of nano-SQUIDS~\cite{Vasyukov13}. In this spirit, atomic-size sensors such as $^{171}$Yb$^{+}$~\cite{Timoney11, Baumgart16, Weidt16} and $^{40}$Ca$^{+}$~\cite{Ruster17} trapped ions, or nitrogen vacancy centers in diamond~\cite{WuJPW16, Santagati19,Haase18} achieve ultimate size-limits for quantum sensors.

Especially interesting is the case of quantum sensors based on $^{171}$Yb$^{+}$ ions that we use as a testbed for our protocol. This ion species encodes the degrees of freedom of the sensor in its $^2 S_{\frac{1}{2}}$ spin manifold whose hyperfine levels present a negligible spontaneous emission rate~\cite{Olmschenk07}. The latter makes the $^{171}$Yb$^{+}$ ion an ideal atomic-size quantum sensor if properly stabilized against decoherence using dynamical decoupling (DD) methods~\cite{Souza12, Biercuk09,Kotler11,CasanovaHW+15,Puebla16,Puebla17,Arrazola18,Arrazola19,Mamin13,Staudacher13,Shi15,Lovchinsky16,Aslam17}. In particular, owing to its resilience against environmental errors and amplitude fluctuations on the microwave (MW) control, the DD scheme leading to the {\it dressed state qubit} has been used for quantum information processing~\cite{Timoney11, Weidt16} and quantum sensing~\cite{Baumgart16}.   Despite this robustness and in close similarity with other sensing techniques, the dressed state qubit approach is restricted to a narrow range in the amplitudes and frequencies of the target electromagnetic signals. A departure from this regime significantly distorts the sensor response and thus makes impossible a direct linear inference of the external field parameters via, e.g., standard fast Fourier transform (FFT) methods.

In this article, we present a method that combines DD techniques to stabilize the quantum sensor with Bayesian inference schemes~\cite{vonderLinden,Gelman}, which enables the accurate estimation of external field parameters from a complex sensor response. This results in a versatile quantum sensing strategy that permits the reconstruction of electromagnetic signals in a wide parameter range, with a minimal previous knowledge of the signal features, and in realistic scenarios involving noise over the sensor and a low number of measurements. As an example, we consider a $^{171}$Yb$^{+}$ ion and demonstrate that Bayesian inference shows a superior performance over standard analysis techniques, such as FFT and least-squares fits. We stress that our method can be adapted to other atomic-size sensors such as $^{40}$Ca$^{+}$ trapped ions or nitrogen vacancy centers in diamond.

\section{Quantum sensor}
We start describing the main features of our quantum sensor device. The $^2 S_{\frac{1}{2}}$ manifold of the $^{171}$Yb$^{+}$ ion comprises four hyperfine levels named $|0\rangle, |\acute 0\rangle, |1\rangle$, and $|-1\rangle$. In an external static magnetic field  $B_z$, the degeneracy of the  $|\acute 0\rangle, |1\rangle , |-1\rangle$ spin levels is removed leading to the diagonal Hamiltonian $H_0= \omega_{\acute 0} |\acute 0\rangle\langle \acute 0| + \sum^1_{j=-1}\omega_{j} |j\rangle\langle j|$, with $\omega_{\pm1} = \frac{A}{4} \pm (\gamma_e-\gamma_n) \frac{B_z}{2}$, 
and $ \omega_{0} =-\omega_{\acute{0}}-A/2= -\frac{3A}{4}  - \frac{(\gamma_e+\gamma_n)^2}{4 A} B_z^2$, where $A=(2\pi) \times 12.643$ GHz~\cite{Olmschenk07} and $\gamma_{e/n}$ is the electronic/nuclear gyromagnetic ratio. We refer to the Supplemental Material (SM) presented in Ref.~\cite{Supplemental}, which includes Refs.~\cite{Griffiths94,Reichenbach07}, for a detailed derivation of $H_0$  and its spectrum. Under a set of control MW drivings, the  $^{171}$Yb$^{+}$ ion Hamiltonian reads
\begin{eqnarray}\label{cleanbigreal}
H &=& H_0 + \mu(t) \ \big(|1\rangle\langle 1| - |-1\rangle\langle -1|\big) \nonumber\\
&+&\sum_j \tilde\Omega_j   \big[ |1\rangle\langle \acute 0| - |1\rangle\langle 0| + |\acute 0\rangle\langle  -1| +  | 0\rangle\langle  -1| + {\rm H.c.} \big],
\end{eqnarray}
where $\tilde\Omega_j = \Omega_j \cos{(\omega_j t +\phi_j)}$ denotes the frequency $\omega_j$, phase $\phi_j$ and Rabi frequency $\Omega_j$ of the $j$th MW driving, and $\mu(t)$ accounts for fluctuations leading to loss of quantum coherence on the magnetically sensitive levels $|1\rangle$ and $|-1\rangle$~\cite{Supplemental}. 

\subsection{Refined atomic-size sensor}

To stabilize the quantum sensor, one has to remove the impact of magnetic field fluctuations from the dynamics, i.e., the term $\mu(t) \ \big(|1\rangle\langle 1| - |-1\rangle\langle -1|\big)$ in Eq.~(\ref{cleanbigreal}). To this end, we tune one of the MW controls in resonance with the $|0\rangle \leftrightarrow |1\rangle$ hyperfine transition, while the other MW-control resonates with $|0\rangle \leftrightarrow |-1\rangle$.  Now, a target electromagnetic field (or signal) can be detected by using either the transition $|\acute0\rangle \leftrightarrow |1 \rangle$ or $|\acute0\rangle \leftrightarrow |-1 \rangle$. Note that a target signal induces the term $\Omega_{\rm tg}\cos(\omega_{\rm tg}t+\phi_{\rm tg})\big[ |1\rangle\langle \acute 0| - |1\rangle\langle 0| + |\acute 0\rangle\langle  -1| +  |0\rangle\langle  -1| + {\rm H.c.} \big]$  in Eq.~\eqref{cleanbigreal}.

\begin{figure}
  \includegraphics[width=1\columnwidth]{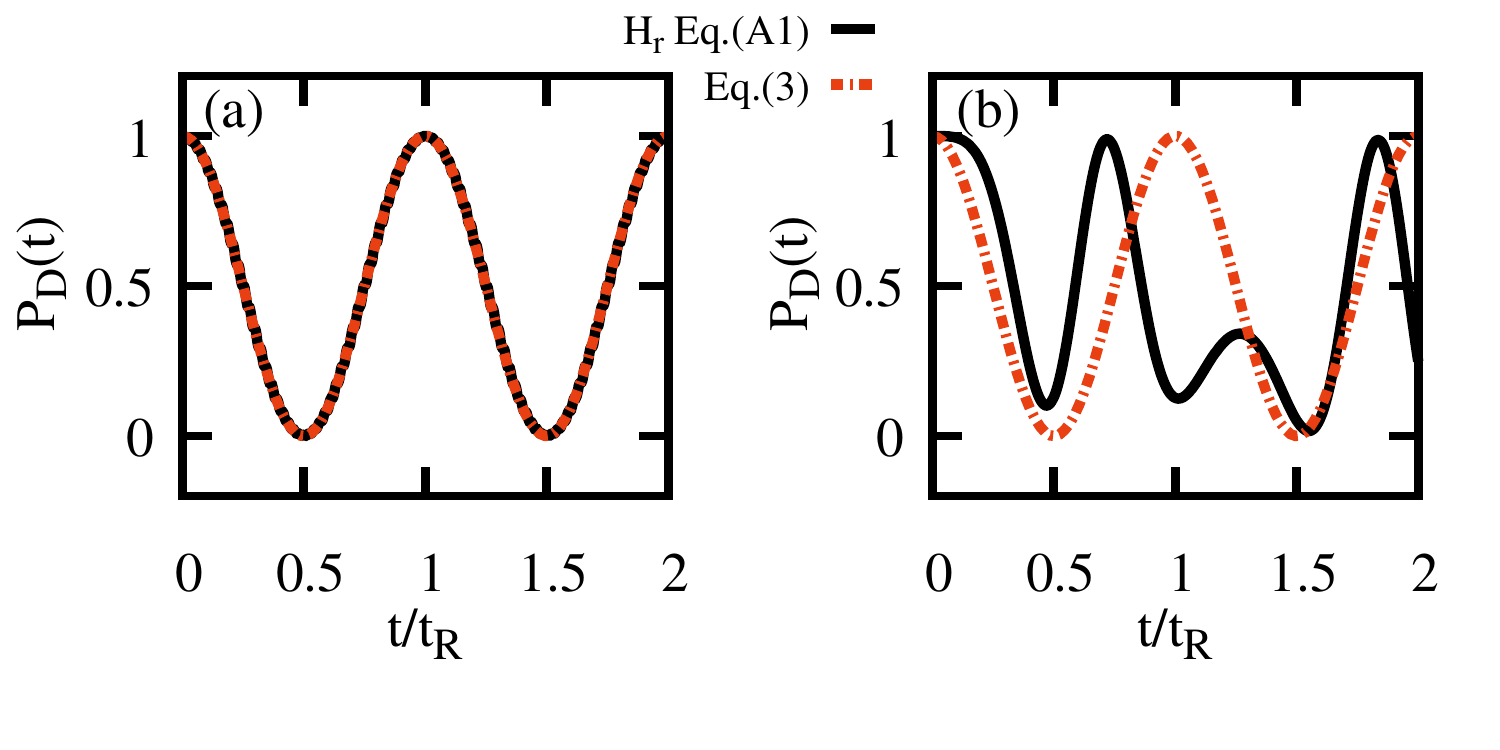}
  \caption{\small{Dynamical behavior of the population $P_D(t)$ with $P_D(0)=1$, $\Omega=2\pi\times 37.27$ kHz as in~\cite{Baumgart16}, and when interacting with a rf-signal with (a) $\Omega_{\rm tg}=2\pi\times 1$ kHz, $\omega_{\rm tg}=2\pi \times 14$ MHz ($B_z=1$ mT), and $\xi=0$, or (b) $\Omega_{\rm tg}=2\pi\times 8$ kHz, $\omega_{\rm tg}=2\pi \times 5.6$ MHz ($B_z=0.4$ mT) with $\xi=2\pi\times 0.25$ kHz. The solid (black) line corresponds to the realistic signal, obtained using $H_{\rm r}$ from Eq.~\eqref{eq:Htot}, while dashed (red) is obtained upon various approximations as shown in Eq.~\eqref{analyticsol} with $t_R=2\pi\sqrt{2}/\Omega_{\rm tg}$. Note the significant deviations with respect to the Rabi oscillations in (b).}}
  \label{fig1}
\end{figure}

The standard procedure to estimate $\Omega_{\rm tg}$ is illustrated in Ref.~\cite{Baumgart16}. This assumes the target field to be on resonance with the $|\acute0\rangle \leftrightarrow |1 \rangle$ transition (that is, $\omega_{\rm tg} = \omega_1 - \omega_{\acute0}$) leading to $H =-\frac{\mu(t)}{\sqrt{2}} (|D\rangle\langle u| + |D\rangle\langle d|+ {\rm H.c.}) + \frac{\Omega}{\sqrt{2}} (|u\rangle\langle u| - |d\rangle\langle d|)
+\frac{\Omega_{\rm tg}}{4} (|u\rangle\langle\acute0| + |d\rangle\langle\acute0| -\sqrt{2}|D\rangle\langle\acute0|+{\rm H.c.})$~\cite{Supplemental}.
The new basis $\{ |u\rangle, |d\rangle, |D\rangle, |\acute{0}\rangle \}$ is $|u\rangle = \frac{1}{\sqrt{2}}(|B\rangle + |0\rangle)$, 
$|d\rangle = \frac{1}{\sqrt{2}}(|B\rangle - |0\rangle)$, $|D\rangle = \frac{1}{\sqrt{2}}(|-1\rangle - |1\rangle)$, $|\acute{0}\rangle = |\acute{0}\rangle$, with $|B\rangle =\frac{1}{\sqrt{2}}(|1\rangle + |-1\rangle)$~\cite{Timoney11, Baumgart16, Weidt16}. The noisy term $-\frac{\mu(t)}{\sqrt{2}} (|D\rangle\langle u| + |D\rangle\langle d|+ {\rm H.c.})$ can be removed since, in the rotating frame defined by the operator $\frac{\Omega}{\sqrt{2}} (|u\rangle\langle u| - |d\rangle\langle d|)$,  it rotates at a speed $\propto \Omega$ which allows one to apply the rotating wave approximation (RWA). Analogously, the terms $-\frac{\mu(t)}{\sqrt{2}} (|D\rangle\langle u| + |D\rangle\langle d|+ {\rm H.c.})$ and $\frac{\Omega_{\rm tg}}{4} (|u\rangle\langle\acute0| + |d\rangle\langle\acute0| +{\rm H.c.})$ average out by invoking the RWA if $\Omega_{\rm tg}\ll \Omega$. One thus finds 
\begin{equation}\label{Rabiosc}
H =-\frac{\Omega_{\rm tg}}{2\sqrt{2}} (|D\rangle\langle\acute0| + |\acute0\rangle\langle D|).
\end{equation}
Eq.~(\ref{Rabiosc}) induces Rabi oscillations between $|D\rangle$ and $ |\acute0\rangle$ at a rate $\propto \Omega_{\rm tg}$. This allows one to find the amplitude of the electromagnetic signal $\Omega_{\rm tg}$ by monitoring, e.g., the population $P_D(t)$ of state $|D\rangle$ at a time $t$. In particular, from Eq.~(\ref{Rabiosc}) and for $P_D(0)=1$, one finds
\begin{equation}\label{analyticsol}
P_D(t)=\cos^2(\pi t/t_R), 
\end{equation}
with $t_R=2\pi\sqrt{2}/\Omega_{\rm tg}$. An example of this purely oscillatory response of the sensor is in Fig.~\ref{fig1}(a). However, a departure from the regime leading to Eq.~(\ref{Rabiosc}) induces significant deviations w.r.t. the periodic behavior predicted by Eq.~(\ref{analyticsol}). An example of such deviations is given in Fig.~\ref{fig1}(b). As we will see later, this challenges the estimation of $\Omega_{\rm tg}$.

A rigorous treatment of the Hamiltonian in Eq.~(\ref{cleanbigreal}) leads to a more involved expression. The resulting Hamiltonian is denoted by $H_{\rm r}$ and reproduced in the Appendix for completeness, Eq.~\eqref{eq:Htot}, while we refer to~\cite{Supplemental} for further details in the derivation of Eq.~\eqref{eq:Htot}. The Hamiltonian $H_{\rm r}$ is our refined model that describes the quantum sensor dynamics in a wide parameter regime. In particular, $H_{\rm r}$ exhibits a non-trivial dependence on $\Omega_{\rm tg}$, as well as on the detuning $\xi$ of the signal w.r.t. the resonant condition, i.e. $\omega_{\rm tg}=\omega_{1}-\omega_{\acute 0}+\xi$. Contrary to Eq.~(\ref{Rabiosc}), $H_{\rm r}$ does not allow us to find analytical expressions for the dynamics of observables such as $P_D(t)$, [cf. Eq.~(\ref{analyticsol})]. However, as we demonstrate later, a specific use of Bayesian methods permits an accurate estimation of target signals in the wide parameter regime described by $H_{\rm r}$ that surpass the performance of standard techniques such as FFT or least-squares methods. 

Regarding the noise sources included in $H_{\rm r}$, we have verified that their effect on the sensor dynamics during the time scales considered in this work is negligible. In this respect, one should note that the scheme in $H_{\rm r}$ includes two MW drivings that eliminate the noise effects induced by, firstly, $\mu(t)$ and, secondly, by Rabi frequency fluctuations. A specific assessment on this  -- including noise sources taken from Ref.~\cite{Baumgart16} -- can be found in~\cite{Supplemental} which includes Refs.~\cite{Uhlenbeck30,Gillespie96a,Gillespie96b,Cai12,Mikelsons15}.

In order to simulate an experimental acquisition of data, we proceed as follows: The data, denoted by ${\bf D}$, is generated by computing the evolution of the quantum sensor state with Hamiltonian $H_{\rm r}$ at different times $t_k$ with $k=1,\ldots, N_p$. The set ${\bf D}$ contains the string of $N_m$ binary outcomes $x_{n;k}\in \{0,1\}$ for each time instant $t_k$ with $n=1,\ldots,N_m$, that is, $x_{n;k}$ are random variables drawn from a Bernoulli distribution $B(P_k)$ where the success probability $P_k$ is obtained from the dynamics of Hamiltonian $H_{\rm r}$. We denote by $X_k=\sum_{n=1}^{N_m}x_{n;k}$ the number of successes recorded at time $t_k$, so that $P^s_k=X_k/N_m$ is the estimation of $P_k$ from ${\bf D}$. In particular, we initialize the system in the state $|D\rangle$ at time $t=0$, and compute the probability $P_k$ of finding it in $|D\rangle$ at time $t_k$, from where the values $x_{n;k}$ are obtained.

\begin{figure}
 \includegraphics[width=1\columnwidth]{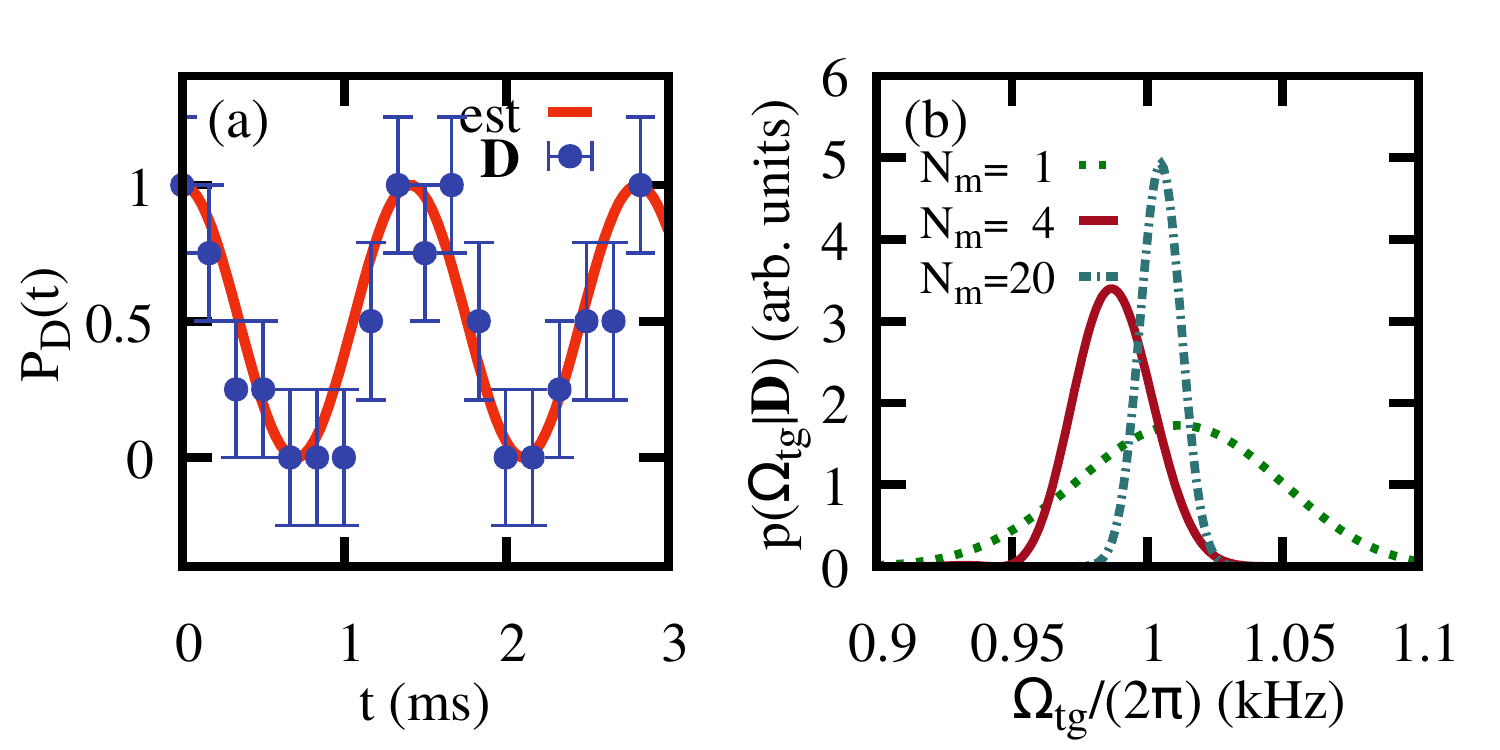}
 \caption{\small{(a) Simulated $N_p=18$ observed populations $P_k^s(t_k)=X_k/N_m$ with $N_m=4$ measurements per point, (points) with error bars indicating a standard deviation due to shot noise~\cite{Supplemental}, together with the reconstructed signal (line) using Bayesian inference for a single unknown parameter $\Omega_{\rm tg}$. (b) Posterior probability distributions for the data in (a) (solid red line) and for the same parameters but $N_m=1$ (dotted green) and $N_m=20$ (dashed blue). See main text for the values of $\Omega_{\rm tg}^{\rm est}$, and Fig.~\ref{fig1}(a) for the rest of parameters.}}
  \label{fig2}
\end{figure}

\section{Bayesian inference and magnetometry}

In the following, we provide the basics of Bayesian inference as relevant to our method (see for example Refs.~\cite{vonderLinden,Gelman} for further details). Let us denote by ${\bf \Theta}=\{\theta_1,\ldots,\theta_M\}$ the set of $M$ unknown parameters which we aim to determine using our quantum sensor from the measured data ${\bf D}$. From Bayes' theorem, the probability $p({\bf \Theta}|{\bf D})\propto p({\bf D}|{\bf \Theta})p({\bf \Theta})$ (typically referred as {\em posterior}) contains the information we can extract from the data given the prior knowledge $p({\bf \Theta})$, and the likelihood $p({\bf D}|{\bf \Theta})$.  The observations $X_k$ that form the data  ${\bf D}$ obey a Bernoulli distribution,  i.e. $p({\bf D}|{\bf \Theta})=\Pi_{k=1}^{N_p}f(X_k,N_m,\tilde{P}_k(t_k;{\bf \Theta}))$, where $f(x,n,p)=n!/(x!(n-x)!) p^x(1-p)^{n-x}$ accounts for the probability of having recorded exactly $x$ successes from $n$ trials drawn from $B(p)$, while   $\tilde{P}_k(t_k;{\bf \Theta})$ denotes the expected probability computed using the Hamiltonian $H_{\rm r}$, given in Eq.~\eqref{eq:Htot}, at time $t_k$ and with parameters ${\bf \Theta}$.  For illustration purposes, we will show the data ${\bf D}$ as $P_k^s(t_k)$ together with the shot-noise uncertainties $\sigma_k$~\cite{Supplemental}.  It is worth remarking that, while magnetic-field and intensity fluctuations have been taken into account to generate the data ${\bf D}$, their effect is negligible in the considered parameter regime~\cite{Supplemental}. For the Bayesian inference, the populations $\tilde{P}_k(t_k;{\bf \Theta})$  are computed without including these noise sources. Having the posterior distribution, one can obtain the estimated mean and variance value of the unknown parameter $\theta_j$ via the marginal distribution $p(\theta_j|{\bf D})$, as $\theta_j^{\rm est}=\int d\theta_j \ \theta_j p(\theta_j|{\bf D})$ and $(\delta\theta_j^{\rm est})^2=\int d\theta_j (\theta_j-\theta_j^{\rm est})^2 p(\theta_j|{\bf D})$, respectively, where the marginal reads as $p(\theta_j|{\bf D})=\int \prod_{i\neq j} d\theta_i \ p({\bf \Theta}|{\bf D})$. 

We exemplify the superior performance of our method over standard analysis techniques with two illustrative cases. For a simplified situation ({\it Case I}) in which Eq.~\eqref{analyticsol} applies leading to a periodic response, we demonstrate Bayesian inference can handle situations with even single shot measurements providing good estimates. %For larger number of measurements per point, least-squares fits provide less accurate results than our method.
When dealing with  more complex signals ({\it Case II}), we show that Bayesian inference from a few number of measurements is able to provide reliable estimates where standard analysis techniques are not applicable in general.

\begin{figure*}
  \includegraphics[width=1.\linewidth]{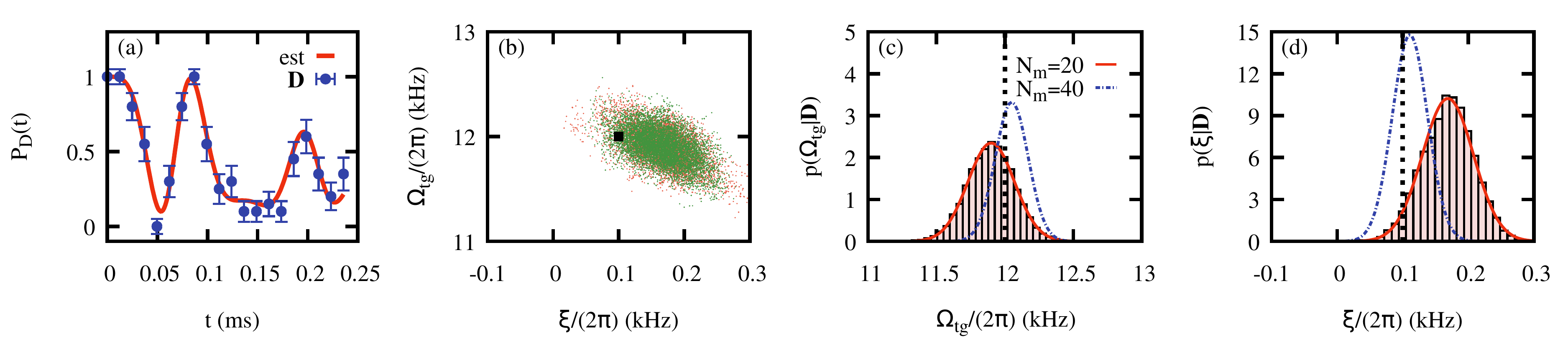}
  \caption{\small{(a) Simulated observations $P_D(t_k)$ with $N_p=N_m=20$ (points) together with the reconstructed signal (line) using Bayesian inference and MCMC for $\Omega_{\rm tg}$ and $\xi$ unknown. The data ${\bf D}$ was generated setting  $B_z=0.5$ mT, with a signal of frequency $\omega_{\rm tg}\approx 2\pi\times 7$ MHz, with  detuning $\xi=2\pi\times 0.1$ kHz and Rabi frequency $\Omega_{\rm tg}=2\pi\times 12$ kHz.  (b) Scatter plot of the recorded values ${\bf \Theta}_j$ during the evolution of the MCMC for two independent chains (red and green dots) with $N_{\rm MC}=10^4$ but removing the burn-in regime. The ideal value is indicated with a black square. Panels (c) and (d) show the histograms for the marginals $p(\Omega_{\rm tg}|{\bf D})$ and $p(\xi|{\bf D})$, respectively, obtained from the MCMC illustrated in (b). The solid red (dashed blue) line corresponds to a Gaussian distribution with equal first and second moments as the marginals for $N_m=20$ ($N_m=40$),  while the dotted black line indicates the ideal value. The estimated values for the case plotted in (a) are $\Omega_{\rm tg}^{\rm est}=2\pi\times 11.90(17)$ kHz and $\xi^{\rm est}=2\pi\times 0.169(39)$ kHz.} }
  \label{fig3}
\end{figure*}

\subsection{Case I}
In this first scenario, ${\bf \Theta}=\{\Omega_{\rm tg}\}$ is the only unknown parameter. Assuming that the RWA can be safely applied and that the target signal is resonant, i.e., $\xi=0$, the sensor is well approximated by Eq.~\eqref{Rabiosc} (cf. Fig.~\ref{fig1}(a))~\cite{Baumgart16}. This allows us to compute the posterior $p(\Omega_{\rm tg}|{\bf D})$ by scanning distinct $\Omega_{\rm tg}$ values, from which $\Omega_{\rm tg}^{\rm est}$ and $\delta\Omega_{\rm tg}^{\rm est}$ can be inferred directly. Here we test our method in the worst case scenario, that  is, when no pre-knowledge about the unknown parameter is available. For that, we consider an uninformative prior, i.e., a flat probability distribution, and an observed signal measured at equally spaced time instances $t_k$ separated by $\Delta t$, such that  $p(\Omega_{\rm tg})\propto 1$ for $0\leq \Omega_{\rm tg}\leq \Omega_{\rm tg}^{\rm max}$, where $\Omega_{\rm tg}^{\rm max}=2\pi/\sqrt{2}\Delta t$. This method can be trivially extended to handle   undersampled or unevenly sampled data~\cite{Supplemental}. 

We simulate an experimental interrogation of the quantum sensor, recording $N_m$ measurements per each of the $N_p$ different time instances. Since $\Delta t\approx 1/6$ ms, it follows $\Omega_{\rm tg}^{\rm max}\approx 2\pi\times 4.2$ kHz. An example is plotted in Fig.~\ref{fig2}(a), together with the estimated signal, while the posterior distributions for different observations are illustrated in Fig.~\ref{fig2}(b). We obtain very precise estimators even with large shot noise, such as the extreme case of single shots (i.e. $N_m=1$). In particular, using same parameters than in Fig.~\ref{fig1}(a) ($\Omega_{\rm tg}=2\pi\times 1$ kHz), we find $\Omega_{\rm tg}^{\rm est}=2\pi \times 1.011(42)$ kHz, $0.988(14)$ kHz and $1.0048(76)$ kHz for three distinct realizations with $N_m=1$, $4$ and $20$ measurements, respectively, where the uncertainty is given by $\delta\Omega_{\rm tg}^{\rm est}$. See~\cite{Supplemental} for further details on the precision of the inferred amplitude $\Omega_{\rm tg}^{\rm est}$ and the string of outcomes for these realizations. In this simple case and for moderate or large number of measurements,  a least-squares fit provide, in average, slightly less accurate results, e.g. $\Omega_{\rm tg}^{\rm est}=2\pi\times 0.947(20)$ kHz for $N_m=4$ (see~\cite{Supplemental} for further realizations and details). As the prior probability distribution is flat, the Bayesian estimators simply correspond to maximum likelihood estimators, which are known to outperform least square fits~\cite{Genschel:10}.
In addition, note that an analysis using standard FFT methods leads to worse estimators. In particular, for the case in Fig.~\ref{fig2}, one obtains $\Omega_{\rm tg}^{\rm est}=2\pi\times 0.94(12)$ kHz~\cite{Supplemental}, which further demonstrates the suitability of Bayesian inference techniques.

\subsection{Case II}
A more realistic situation needs to account for potential non-resonant radiation as well as off-resonant transitions within the quantum sensor. Thus, ${\bf \Theta}=\{\Omega_{\rm tg},\xi\}$ where $\xi$ denotes a detuning w.r.t. the resonant condition, and Eq.~\eqref{eq:Htot} is required (cf. Fig.~\ref{fig1}(b)).  In addition, Markov chain Monte Carlo (MCMC) methods will be employed to efficiently  sample the posterior $p({\bf \Theta}|{\bf D})$~\cite{vonderLinden,Gilks}. For that, we consider independent priors, namely, $p(\Omega_{\rm tg},\xi)=p(\Omega_{\rm tg})p(\xi)$, taking again $p(\Omega_{\rm tg})$ completely uninformative in the region $0\leq \Omega_{\rm tg}\leq 2\pi\times 50$ kHz, while $p(\xi)=\mathcal{N}(0,\sigma_{\xi}^2)$ with $\sigma_{\xi}=2\pi\times 0.25$ kHz, as we expect close to resonant rf-fields. By randomly choosing an initial point ${\bf \Theta}_0$ from the prior,  we rely on a standard Metropolis algorithm to sample the posterior~\cite{Gilks}. After $j$ steps, the proposed point ${\bf \Theta}_{j+1}$ obtained from $\mathcal{N}({\bf \Theta}_j,\tilde{{\bf \sigma}}_p^2)$ where $\tilde{{\bf \sigma}}_p^2=\{\tilde{\sigma}^2_{\Omega},\tilde{\sigma}_{\xi}^2\}$ refers to the variance in the proposal distributions, is accepted with probability $\alpha=\min(1,p({\bf \Theta}_{j+1}|{\bf D})/p({\bf \Theta}_j|{\bf D}))$. After a sufficient number of steps, $N_{\rm MC}\gg 1$, the recorded ${\bf \Theta}$ values provide an accurate sampling of $p({\bf \Theta}|{\bf D})$ and the marginals can be easily computed. Convergence of the MCMC can be checked by the mixing of different Markov chains~\cite{Gilks,Supplemental}. Although we illustrate the working method for this case of study with a single example, we stress that the following procedure is general and can be applied to different situations. 

In Fig.~\ref{fig3} we have considered a set of data {\bf D} obtained for a rf-signal with $\omega_{\rm tg}\approx 2\pi\times 7$ MHz ($B_z=0.5$ mT), a detuning of $\xi=2\pi\times 0.1$ kHz and amplitude $\Omega_{\rm tg}=2\pi\times 12$ kHz, and $\Omega=2\pi\times 37.27$ kHz that protects the sensor against magnetic-field fluctuations, while the data has been generated with $N_p=N_m=20$ (cf. Figs.~\ref{fig3}(a)). For the MCMC we observe that $\tilde{\sigma}_{\Omega}=10\tilde{\sigma}_{\xi}=2\pi\times 0.1$ kHz yields a good mixing (cf. Fig.~\ref{fig3}(b) and~\cite{Supplemental}), so that the effective size of the MCMC (number of accepted points) amounts approximately to $N_{\rm MC}/2$ (cf. Fig.~\ref{fig3}(b)). We remove the first $200$ steps to avoid the burn-in regime~\cite{vonderLinden,Gilks}. In Figs.~\ref{fig3}(c) and (d) we show the marginals $p(\Omega_{\rm tg}|{\bf D})$ and  $p(\xi|{\bf D})$, respectively, obtained upon $N_{\rm MC}=10^4$ steps for five independent Markov chains, which lead to $\Omega_{\rm tg}^{\rm est}=2\pi\times 11.90(17)$ kHz and $\xi^{\rm est}=2\pi\times 0.169(39)$ kHz, very close to the ideal values.   

The complex and non-harmonic response of the sensor challenges the determination of the unknown parameters for single shot acquisitions~\cite{Supplemental}. However, for a reduced number of measurements per point, e.g. $N_m=4$, we still find good estimates for the amplitude $\Omega_{\rm tg}^{\rm est}=2\pi\times 12.91(44)$ kHz, although the data may be better explained under distinct detunings, $\xi^{\rm est}=2\pi\times -0.112(90)$ kHz.  In a similar manner, by reducing the shot-noise, more accurate estimates can be obtained, e.g. $\Omega_{\rm tg}^{\rm est}=2\pi\times 12.05(12)$ kHz and $\xi^{\rm est}=2\pi\times 0.111(27)$ kHz for $N_m=40$ measurements per point (cf. Figs.~\ref{fig3}(c) and (d)). We provide the string of outcomes ${\bf D}$ for each of the realizations and more examples in~\cite{Supplemental}. Finally, it is worth mentioning that neither least-squares nor FFT techniques are useful in this case due to the complex signal structure. As illustrated in~\cite{Supplemental}, a non-linear least-squares fit to the dynamics dictated by $H_{\rm r}$ is unable to find suitable parameters unless initialized close to the ideal values and unsuitable to tackle more complex cases such as in bi-modal posterior distributions~\cite{Supplemental}, while at the same time FFT methods exhibit an intricate frequency spectrum of the data ${\bf D}$ hindering the identification of the unknown parameters.

\section{Conclusions}

We presented a protocol relying on Bayesian methods that enhance significantly the performance of quantum sensors in realistic scenarios. In particular, we have demonstrated that  a quantum sensor can be used even when the character of target signals, as well as the presence of noise and a reduced number of measurements, spoil its ideal functioning leading to strong deviations of the sensor from a simple harmonic response. We illustrate this scheme using a ${}^{171}{\rm Yb}^+$ trapped-ion, and relying on standard MCMC methods if so required by the parameter regime. Our results showcase the suitability of Bayesian inference with respect to standard analysis techniques for parameter estimation. Our method therefore paves the way to use quantum sensors under realistic conditions, significantly extending their working region and reducing the detection times, thus enhancing their adaptability to different scenarios.

\begin{acknowledgments}
  We thank Benjamin D'Anjou for helpful comments, and acknowledge financial support from Spanish Government via PGC2018-095113-B-I00 (MCIU/AEI/FEDER, UE), Basque Government via IT986-16, as well as from QMiCS (820505) and OpenSuperQ (820363) of the EU Flagship on Quantum Technologies, and the EU FET Open Grant Quromorphic. J.C. acknowledges the Ram{\'o}n y Cajal program (RYC2018- 025197-I) and support from the UPV/EHU through the grant EHUrOPE. M. B. P. acknowledges support by the ERC Synergy Grant HyperQ, the EU Flagship project AsteriQs and the BMBF projects Nanospin and DiaPol.  J. F. H. acknowledges support by the Alexander von Humboldt Foundation in form of a Feodor-Lynen Fellowship. R. P. and M. P. acknowledge the support by the SFI-DfE Investigator Programme (grant 15/IA/2864). M. P. acknowledges the H2020 Collaborative Project TEQ (Grant Agreement 766900), the Leverhulme Trust Research Project Grant UltraQuTe (grant RGP-2018-266), the Royal Society Wolfson Fellowship (RSWF/R3/183013) and the UK EPSRC (grant EP/T028106/1). 
\end{acknowledgments}

\appendix
\section{Refined model for the atomic-size sensor}
A more rigorous treatment of the Hamiltonian given in Eq.~\eqref{cleanbigreal} can be written in the basis $\{ |u\rangle, |d\rangle, |D\rangle, |\acute{0}\rangle \}$ as
\begin{widetext}
  \begin{eqnarray}\label{eq:Htot}
H_{\rm r}&=&\frac{-\mu(t)}{\sqrt{2}} (|D\rangle\langle u| +  |D\rangle\langle d| + {\rm H.c.}) + \frac{\Omega}{\sqrt{2}}(|u\rangle\langle u| -  |d\rangle\langle d|)-   \bigg[\frac{\Omega}{2\sqrt{2}}(|u\rangle\langle u| -  |d\rangle\langle d|) +  \frac{\Omega}{4}  (|u\rangle\langle D| +  |D\rangle\langle d|) \nonumber\\
&&-  \frac{\Omega}{4}  (|D\rangle\langle u| +  |d\rangle\langle D|)  \bigg] e^{i\gamma_e B_z t} +\bigg[\frac{\Omega_{\rm tg}}{4} (|u\rangle\langle\acute0| + |d\rangle\langle\acute0|) - \frac{\Omega_{\rm tg}}{2\sqrt{2}} |D\rangle\langle\acute0|\bigg]e^{-i\xi t} + {\rm H.c.}\nonumber\\
&+&\frac{\Omega_{\rm tg}}{2} \bigg(\frac{1}{2} |u\rangle\langle \acute0| + \frac{1}{2} |d\rangle\langle \acute0| - \frac{1}{\sqrt{2}} |D\rangle\langle \acute0|  \bigg) e^{2i(\frac{\gamma_e B_z}{2} - \frac{\gamma_e^2}{4A}B_z^2)t}e^{i\xi t} + \frac{\Omega_{\rm tg}}{2}\bigg(\frac{1}{2}|\acute 0\rangle\langle u| +  \frac{1}{2}|\acute 0\rangle\langle d| + \frac{1}{\sqrt 2}|\acute 0\rangle\langle D| \bigg)  e^{i \gamma_e B_z t}e^{ i\xi t}  + {\rm H.c.}\nonumber\\
&+&\frac{\Omega_{\rm tg}}{2}\bigg(\frac{1}{2}|\acute 0\rangle\langle u| +  \frac{1}{2}|\acute 0\rangle\langle d| + \frac{1}{\sqrt 2}|\acute 0\rangle\langle D| \bigg)e^{i \frac{\gamma^2_e}{2A} B^2_z t}e^{ - i\xi t} + {\rm H.c.}
  \end{eqnarray}
  \end{widetext}
  where the first term accounts for magnetic-field fluctuations in the states $ | D\rangle$ and $| u\rangle $. The rest of the terms appear due to both, a non-resonant target signal $\omega_{\rm tg}=\omega_{1}-\omega_{\acute 0}+\xi$ with a detuning $\xi$, as well as a large Rabi frequency $\Omega_{\rm tg}$ compared to the frequency $\omega_{\rm tg}$ of the rf-signal. The previous Hamiltonian includes two MW controls with amplitudes $\Omega$. See~\cite{Supplemental} for the details of the derivation.

\widetext
\clearpage
\begin{center}
\textbf{\large Supplemental Material\\Versatile Atomic Magnetometry Assisted by Bayesian Inference}
\end{center}
%%%%%%%%%% Merge with supplemental materials %%%%%%%%%%
%%%%%%%%%% Prefix a "S" to all equations, figures, tables and reset the counter %%%%%%%%%%
\setcounter{equation}{0}
\setcounter{figure}{0}
\setcounter{table}{0}
\setcounter{page}{1}
\makeatletter
\renewcommand{\theequation}{S\arabic{equation}}
\renewcommand{\thefigure}{S\arabic{figure}}
\renewcommand{\bibnumfmt}[1]{[S#1]}
\renewcommand{\citenumfont}[1]{S#1}

\begin{center}
  R. Puebla,${}^{1,2}$ Y. Ban,${}^{3,4}$ J. F. Haase,${}^{5,6}$ M. B. Plenio,${}^7$ M. Paternostro,${}^2$ and J. Casanova${}^{3,8}$\\
  \vspace{0.2cm}{\small $^1${\em Instituto de F{\'i}sica Fundamental, IFF-CSIC, Calle Serrano 113b, 28006 Madrid, Spain}\\
    ${}^2${\em Centre for Theoretical Atomic, Molecular, and Optical Physics,\\ School of Mathematics and Physics, Queen's University, Belfast BT7 1NN, United Kingdom}\\
    ${}^3${\em Department of Physical Chemistry, University of the Basque Country UPV/EHU, Apartado 644, 48080 Bilbao, Spain}\\
    ${}^4${\em School of Materials Science and Engineering, Shanghai University, 200444 Shanghai, China}\\
    ${}^5${\em Institute for Quantum Computing, University of Waterloo, Waterloo, Ontario, Canada, N2L 3G1}\\
    ${}^6${\em Department of Physics \& Astronomy, University of Waterloo, Waterloo, Ontario, Canada, N2L 3G1}\\
  $^7${\em Institute of Theoretical Physics and IQST, Albert-Einstein Allee 11, Universit\"{a}t Ulm,
      89069 Ulm, Germany}\\
    ${}^8${\em IKERBASQUE, Basque  Foundation  for  Science,  Maria  Diaz  de  Haro  3,  48013  Bilbao, Spain}}
  \end{center}

\section*{I. $^{171}$Yb$^{+}$ sensor energy levels}\label{app:yb}
In this appendix we provide a summary of  the $^{171}$Yb$^{+}$  physical properties. We are interested in the long-lived  $^{2}S_{\frac{1}{2}}$ manifold of the $^{171}$Yb$^{+}$ ion~\cite{Olmschenk07SM}. This means $L=0$, i.e. zero angular momentum, and spin $S=\frac{1}{2}$ according to the general spectroscopic notation  $^{2S+1} L_J $.   The hyperfine interaction in this manifold is created because the $^{171}$Yb$^+$ nucleus carries  a spin $I=\frac{1}{2}$ which interacts with the electronic spin~\cite{Olmschenk07SM} leading to the following Hamiltonian
\begin{equation}\label{hyperfinesimple}
H = A \ {\bf J} \cdot {\bf I},
\end{equation}
where ${\bf J}$ is a spin-1/2 operator for the electron (note we are in the $^{2}S_{\frac{1}{2}}$ manifold), and ${\bf I}$ is a nuclear spin-1/2 operator. This means that we can write ${\bf I} = \frac{1}{2}\vec{\sigma}_1$ and ${\bf J} = \frac{1}{2}\vec{\sigma}_2$ where $\vec{\sigma}_{1,2} = (\sigma^x_{1, 2}, \sigma^y_{1,2}, \sigma^z_{1,2})$. In addition, $A$ is the magnetic hyperfine constant which is $A \approx (2\pi) \times 12.643$ GHz as measured in~\cite{Olmschenk07SM}. The Hamiltonian that describes this situation once a magnetic field $\vec{B}=B_z \hat{z}$ is included reads 
\begin{equation}\label{Shyperfine}
H = A \ {\bf J} \cdot {\bf I} + g_J \mu_B \ {\bf J}  \cdot \vec{B}  - g_I \mu_N  \ {\bf I} \cdot \vec{B}
\end{equation}
where $g_J = \bigg[1 + (g_S -1) \frac{j(j+1) - l(l+1) + s(s+1)}{2 j(j+1)} \bigg]$  is the Land\'e $g$-factor of the atom (see for example~\cite{Griffiths94SM}) and $g_S \approx 2.0023$ is the responsible of the anomalous gyromagnetic factor of the electron spin (in our case $j\!=\!s\!=1/2$ and $l\!=\!0$, hence $g_J = 1+(g_S -1)=g_S$). Note that a similar expression for the $^{1}P_1$ subspace can be found in~\cite{Reichenbach07SM}).

The static magnetic field leads to a Zeeman splitting of the energy levels. If we redefine $g_S \mu_B\equiv \hbar \gamma_e$ and $g_I \mu_N\equiv \hbar \gamma_n$, where $\gamma_e = (2\pi)\times 2.8024$ MHz/G and $\gamma_n$ the gyromagnetic factor of the $^{171}{\rm Yb}^{+}$ nucleus, with $\gamma_n\equiv \gamma_{^{171}\rm Yb^{+}} = (2\pi)\times 4.7248$ kHz/G, i.e.  $ \gamma_n \ll  \gamma_e $, the Hamiltonian~(\ref{Shyperfine}) can be written as 
\begin{equation}\label{simple}
H = A \ {\bf J} \cdot {\bf I}  + \gamma_e B_z J_z - \gamma_n B_z I_z.
\end{equation}
In the basis $\{ |1 1\rangle, |1 0\rangle, |0 1\rangle, |0 0\rangle\}$ (with $\sigma_z|1\rangle = |1\rangle$ and 
$\sigma_z|0\rangle = -|0\rangle$) one can write
\begin{equation}\label{Smatrix}
H=\left(\begin{array}{cccc}
\frac{A}{4} + (\gamma_e-\gamma_n) \frac{B_z}{2} &0&0&0\\
0&-\frac{A}{4} + (\gamma_e+\gamma_n) \frac{B_z}{2}&\frac{A}{2}&0\\
0&\frac{A}{2}&-\frac{A}{4} - (\gamma_e+\gamma_n) \frac{B_z}{2}&0\\
0&0&0&\frac{A}{4} - (\gamma_e-\gamma_n) \frac{B_z}{2}
\end{array}\right).
\end{equation}
The states $ |1\rangle  = |1 1\rangle$ and $|-1\rangle = |0 0\rangle$, have the eigenfrequencies $\omega_{1} = \frac{A}{4} + (\gamma_e-\gamma_n) \frac{B_z}{2}$ and $\omega_{-1} = \frac{A}{4} - (\gamma_e-\gamma_n) \frac{B_z}{2}$, while  diagonalization of Eq.~(\ref{Smatrix}) leads to two additional energies, namely,  $\omega_{\acute0} = -\frac{A}{4} + \frac{A}{2}\sqrt{1 + \left[\frac{(\gamma_e+\gamma_n)B_z}{A}\right]^2}$ and $\omega_{0} = -\frac{A}{4} - \frac{A}{2}\sqrt{1 + \left[\frac{(\gamma_e+\gamma_n)B_z}{A}\right]^2}$. The latter expressions  can be expanded if $\left[\frac{(\gamma_e+\gamma_n)B_z}{A}\right]\ll 1$ (note this is our case since we consider low values for $B_z$) leading to $\omega_{\acute{0}} \approx  \frac{A}{4}  + \frac{(\gamma_e+\gamma_n)^2}{4 A}B_z^2$ and  $\omega_{0} \approx - \frac{3A}{4}  - \frac{(\gamma_e+\gamma_n)^2}{4 A} B_z^2$. The quantity $\omega_{\acute0} - \omega_{0} =  A + \frac{(\gamma_e+\gamma_n)^2}{2 A}B_z^2$  where the factor $ \frac{(\gamma_e+\gamma_n)^2}{2 A}\approx (2\pi)\times310.8 \frac{\rm Hz}{\rm G^2}$ is known as the second-order Zeeman shift~\cite{Olmschenk07SM}.  

As a summary, Hamiltonian~(\ref{Smatrix}) has the following eigenstates and eigenvalues 
\begin{eqnarray}\label{summary}
|1\rangle  = |1 1\rangle &\longrightarrow& \omega_{1} = \frac{A}{4} + (\gamma_e-\gamma_n) \frac{B_z}{2}\nonumber\\
|-1\rangle  = |0 0\rangle &\longrightarrow& \omega_{-1} = \frac{A}{4} - (\gamma_e-\gamma_n) \frac{B_z}{2}\nonumber\\
|\acute{0}\rangle = \alpha |1 0\rangle + \beta |0 1\rangle  &\longrightarrow& \omega_{\acute{0}} \approx \frac{A}{4}  + \frac{(\gamma_e+\gamma_n)^2}{4 A} B_z^2\nonumber\\
|0\rangle = \gamma |1 0\rangle + \delta |0 1\rangle  &\longrightarrow& \omega_{0} \approx -\frac{3A}{4}  - \frac{(\gamma_e+\gamma_n)^2}{4 A} B_z^2
\end{eqnarray}
where $\alpha = \sqrt{\frac{1}{1 + \left(\frac{\omega_{\acute0}-a}{b}\right)^2}}$, $\beta = \frac{\omega_{\acute0}-a}{b} \sqrt{\frac{1}{1 + \left(\frac{\omega_{\acute0}-a}{b}\right)^2}}$, $\gamma = \sqrt{\frac{1}{1 + \left(\frac{\omega_{0}-a}{b}\right)^2}}$, and $\delta = \frac{\omega_{0}-a}{b} \sqrt{\frac{1}{1 + \left(\frac{\omega_{0}-a}{b}\right)^2}}$, with $a = -\frac{A}{4} + (\gamma_e+\gamma_n) \frac{B_z}{2}$ and $b = \frac{A}{2}$.

\begin{figure}[t]
\includegraphics[width=0.9\columnwidth]{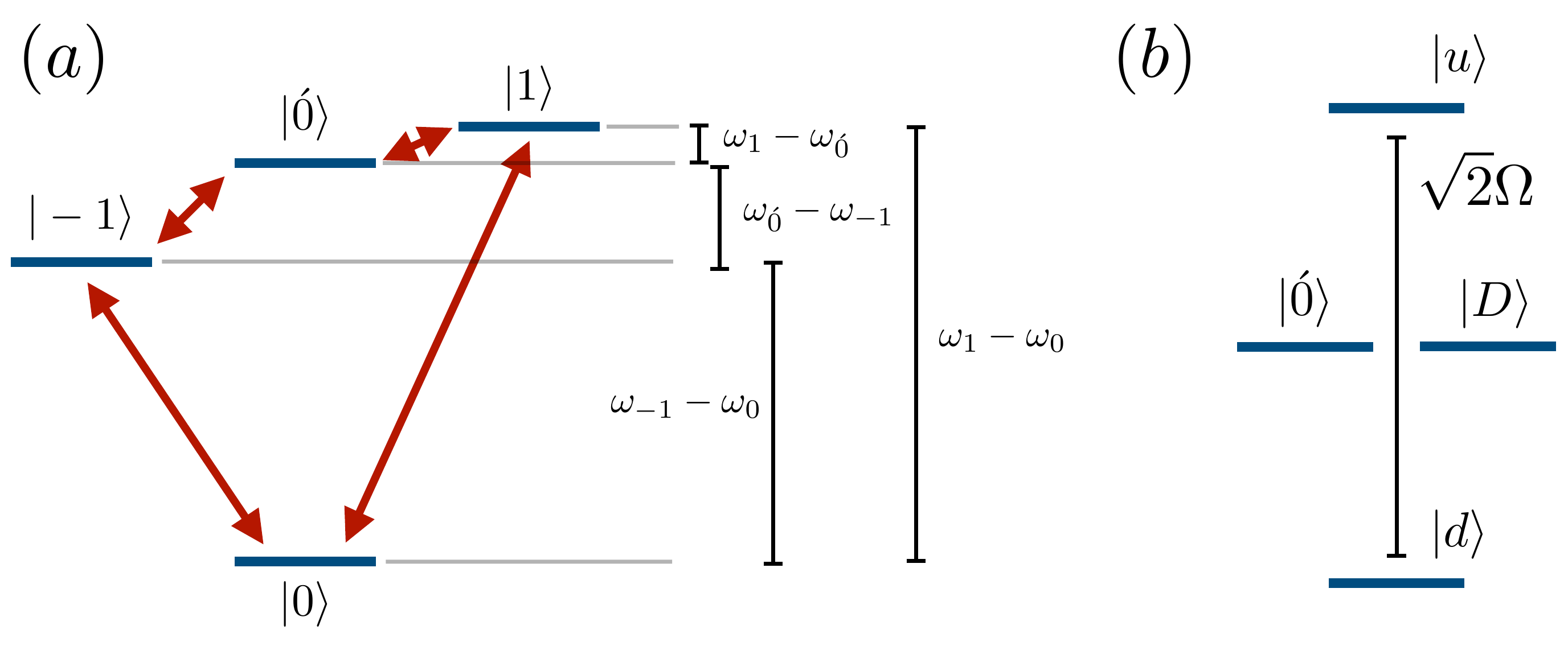}\caption{(a) Energy scheme of the  $^{171}{\rm Yb}^{+}$ atom under the presence of a magnetic field according to Eq.~(\ref{energydiffs}) as well as the allowed transitions that appear as a consequence of introducing an external MW field. (b) Dressed state qubit basis and the corresponding energy differences (cf. Eqs.~\eqref{idealcase} and~\eqref{SHRabi}). \label{Sdrseedqubit} } 
\end{figure}

In this new basis the Hamiltonian~(\ref{simple}) can be written as
\begin{equation}\label{energydiffs}
H = \omega_1 |1\rangle\langle 1| + \omega_{\acute 0} |\acute 0\rangle\langle \acute 0| + \omega_{-1} |-1\rangle\langle -1| + \omega_{0} | 0\rangle\langle 0|.
\end{equation}
In Fig.~\ref{Sdrseedqubit}(a) we have sketched the energy diagram of the $^{171}$Yb$^+$ ion's $^{2}S_{\frac{1}{2}}$ manifold. 
We can induce transitions among the states in the diagonal basis $\{ |0\rangle, |-1 \rangle, |\acute 0 \rangle, |1\rangle \}$ with radiofrequency and microwave fields. For example, the driving $B_x \cos{(\omega t + \phi)}$ leads to the following interaction 
\begin{equation}
H = A \ {\bf J} \cdot {\bf I}  + \gamma_e B_z J_z - \gamma_n B_z I_z + \gamma_e B_x  J_x \cos{(\omega t +\phi)} - \gamma_n B_x I_x \cos{(\omega t + \phi)}.
\end{equation}
Or, in the diagonal basis 
\begin{equation}
H = \omega_1 |1\rangle\langle 1| + \omega_{\acute 0} |\acute 0\rangle\langle \acute 0| + \omega_{-1} |-1\rangle\langle -1| + \omega_{0} | 0\rangle\langle 0| + \gamma_e B_x  J_x \cos{(\omega t +\phi)} - \gamma_n B_x I_x \cos{(\omega t+\phi)}.
\end{equation}
To see the induced transitions as a consequence of the newly introduced driving field, we have to expand $J_x$ and $I_x$ in the new basis. With the help of the expressions 
\begin{eqnarray}
|1 0\rangle &=& \frac{\delta}{\delta \alpha - \beta \gamma} \bigg(|\acute0 \rangle - \frac{\beta}{\delta}  | 0 \rangle \bigg),\nonumber\\
|0 1\rangle &=& \frac{\gamma}{\gamma \beta - \alpha \delta} \bigg(|\acute0 \rangle - \frac{\alpha}{\gamma}  | 0 \rangle \bigg),
\end{eqnarray}
one can easily find  
\begin{equation}
J_x = \frac{1}{2(\gamma \beta - \alpha \delta)} \bigg[ \gamma  |1\rangle\langle \acute{0}| -  \alpha  |1\rangle\langle 0|  -\delta  |\acute{0}\rangle\langle-1|  + \beta |0\rangle\langle -1|  + \rm{H.c.} \bigg],
\end{equation}
and 
\begin{equation}
I_x = \frac{1}{2(\gamma \beta - \alpha \delta)} \bigg[ -\delta   |1\rangle\langle \acute{0}| +  \beta |1\rangle\langle 0| + \gamma  |\acute{0}\rangle\langle-1| - \alpha |0\rangle\langle -1|  + \rm{H.c.} \bigg].
\end{equation}
In this manner, one can  write 
\begin{eqnarray}\label{Sclean}
H &=& \omega_1 |1\rangle\langle 1| + \omega_{\acute 0} |\acute 0\rangle\langle \acute 0| + \omega_{-1} |-1\rangle\langle -1| + \omega_{0} | 0\rangle\langle 0| \nonumber\\
&+& \frac{B_x}{2}  \bigg[ c_{1 \acute0} |1\rangle\langle \acute{0}| +  c_{1 0} |1\rangle\langle 0| + c_{\acute 0 -1} |\acute0\rangle\langle -1| + c_{ 0 -1} |0\rangle\langle -1|  + {\rm H.c.} \bigg]   \cos{(\omega t +\phi)} 
\end{eqnarray}
where 
\begin{eqnarray}\label{Scoefficients}
c_{1 \acute{0}}=  \frac{1}{\gamma \beta - \alpha \delta}  (\gamma_e \gamma  + \gamma_n \delta),\nonumber\\
c_{1 0}=  \frac{-1}{\gamma \beta - \alpha \delta}  (\gamma_e \alpha  + \gamma_n \beta),\nonumber\\
c_{\acute{0} -1}=  \frac{-1}{\gamma \beta - \alpha \delta}  (\gamma_e \delta  + \gamma_n \gamma),\nonumber\\
c_{0 -1}=  \frac{1}{\gamma \beta - \alpha \delta}  (\gamma_e \beta  + \gamma_n \alpha).\nonumber\\
\end{eqnarray}
In  Hamiltonian~(\ref{Sclean}) we can see the allowed transitions that would occur when the frequency $\omega$ of the external driving is on resonance with the corresponding  energy difference of each of the transitions. More specifically, in the rotating frame of  $H_0 =  \omega_1 |1\rangle\langle 1| + \omega_{\acute 0} |\acute 0\rangle\langle \acute 0| + \omega_{-1} |-1\rangle\langle -1| + \omega_{0} | 0\rangle\langle 0|$, one can find 
\begin{equation}
H = \frac{B_x}{4}\bigg[ c_{1 \acute0} |1\rangle\langle \acute{0}| e^{i(\omega_1 -  \omega_{\acute 0})t} +  c_{1 0} |1\rangle\langle 0| e^{i(\omega_1 -  \omega_{0})t} + c_{\acute 0 -1} |\acute0\rangle\langle -1| e^{i(\omega_{\acute{0}} -  \omega_{-1})t} + c_{ 0 -1} |0\rangle\langle -1| e^{i(\omega_0 -  \omega_{-1})t}  + {\rm H.c.} \bigg]  \bigg[e^{i(\omega t + \phi)} + e^{-i(\omega t + \phi)}\bigg].
\end{equation}

The required energy of each transition reads as (in Fig.~\ref{Sdrseedqubit} a) one can see the energy diagram)
\begin{eqnarray}\label{Stransitions}
\omega_{1} - \omega_{\acute0} &\approx& \frac{\gamma_e B_z}{2} - \frac{\gamma_e^2}{4A}B_z^2,\nonumber\\
\omega_{1} - \omega_{0}&\approx& A + \frac{\gamma_e B_z}{2} + \frac{\gamma_e^2}{4A}B_z^2,\nonumber\\
\omega_{\acute{0}} -  \omega_{-1}&\approx& \frac{\gamma_e B_z}{2} + \frac{\gamma_e^2}{4A}B_z^2,\nonumber\\
\omega_{-1} -  \omega_{0}&\approx& A - \frac{\gamma_e B_z}{2} + \frac{\gamma_e^2}{4A}B_z^2.
\end{eqnarray}
In addition, when several drivings act on the system, one can straightforwardly extend Hamiltonian~(\ref{Sclean}) to  
\begin{eqnarray}\label{cleanbig}
H &=& \omega_1 |1\rangle\langle 1| + \omega_{\acute 0} |\acute 0\rangle\langle \acute 0| + \omega_{-1} |-1\rangle\langle -1| + \omega_{0} | 0\rangle\langle 0| \nonumber\\
&+&\sum_j \frac{B_x^j}{2}  \bigg[ c_{1 \acute0} |1\rangle\langle \acute{0}| +  c_{1 0} |1\rangle\langle 0| + c_{\acute 0 -1} |\acute0\rangle\langle -1| + c_{ 0 -1} |0\rangle\langle -1|  + {\rm H.c.} \bigg]   \cos{(\omega_j t +\phi_j)}. 
\end{eqnarray}
In order to complete the model, we have to consider the effect of magnetic-field fluctuations. Hence, we introduce a noise source in $\omega_1$ and $\omega_{-1}$. In this respect, note that the $|0\rangle$ and $|\acute0\rangle$ hyperfine levels also fluctuate but with a much more smaller intensity since the magnetic field enters trough the small  second-order Zeeman shift, cf. Eq.~\eqref{summary}. This leads to  
\begin{eqnarray}\label{Scleanbigreal}
H &=& \omega_1 |1\rangle\langle 1| + \omega_{\acute 0} |\acute 0\rangle\langle \acute 0| + \omega_{-1} |-1\rangle\langle -1| + \omega_{0} | 0\rangle\langle 0| + \mu(t) (|1\rangle\langle 1| - |-1\rangle\langle -1|)\nonumber\\
&+&\sum_j \frac{B_x^j}{2}  \bigg[ c_{1 \acute0} |1\rangle\langle \acute{0}| +  c_{1 0} |1\rangle\langle 0| + c_{\acute 0 -1} |\acute0\rangle\langle -1| + c_{ 0 -1} |0\rangle\langle -1|  + {\rm H.c.} \bigg]   \cos{(\omega_j t +\phi_j)}. 
\end{eqnarray}
Note that the noisy term $\mu(t) (|1\rangle\langle 1| - |-1\rangle\langle -1|)$ can be understood by inspecting the expressions for $\omega_1$ and $\omega_{-1}$ and considering that $B_z$ carries a fluctuation such that the static magnetic field equals to $B_z[1 + \xi(t)]$, and  $B_z\xi(t)(\gamma_e-\gamma_n)/2 = \mu(t)$.

\section*{II. Derivation of Eqs. (2) and (A1) of main text}
Here we provide the details to derive Eqs. (2), and thus (3), and the Hamiltonian $H_{\rm r}$ given in Eq. (A1) of the main text. In particular, in {\bf Section~II A} we find Eq. (2) that describes the effective Hamiltonian used in standard measurements schemes. In {\bf Section~II B} we derive Eq. (A1) which is the target Hamiltonian $H_{\rm r}$ we use in the main text. Equation (A1) is reproduced here for convenience
\begin{eqnarray}
  H_{\rm r}&=&\frac{-\mu(t)}{\sqrt{2}} (|D\rangle\langle u| +  |D\rangle\langle d| + {\rm H.c.}) + \frac{\Omega}{\sqrt{2}}(|u\rangle\langle u| -  |d\rangle\langle d|)\nonumber\\
  &-&  \bigg[\frac{\Omega}{2\sqrt{2}}(|u\rangle\langle u| -  |d\rangle\langle d|) +  \frac{\Omega}{4}  (|u\rangle\langle D| +  |D\rangle\langle d|) -  \frac{\Omega}{4}  (|D\rangle\langle u| +  |d\rangle\langle D|)  \bigg] e^{i\gamma_e B_z t} + {\rm H.c.}\nonumber\\
&+&\bigg[\frac{\Omega_{\rm tg}}{4} (|u\rangle\langle\acute0| + |d\rangle\langle\acute0|) - \frac{\Omega_{\rm tg}}{2\sqrt{2}} |D\rangle\langle\acute0|\bigg]e^{-i\xi t} + {\rm H.c.}\nonumber\\
&+&\frac{\Omega_{\rm tg}}{2} \bigg(\frac{1}{2} |u\rangle\langle \acute0| + \frac{1}{2} |d\rangle\langle \acute0| - \frac{1}{\sqrt{2}} |D\rangle\langle \acute0|  \bigg) e^{2i(\frac{\gamma_e B_z}{2} - \frac{\gamma_e^2}{4A}B_z^2)t}e^{i\xi t} + {\rm H.c.}\nonumber\\
&+&\frac{\Omega_{\rm tg}}{2}\bigg(\frac{1}{2}|\acute 0\rangle\langle u| +  \frac{1}{2}|\acute 0\rangle\langle d| + \frac{1}{\sqrt 2}|\acute 0\rangle\langle D| \bigg)  e^{i \gamma_e B_z t}e^{ i\xi t}  + {\rm H.c.}\nonumber\\ \label{eq:Htot}
&+&\frac{\Omega_{\rm tg}}{2}\bigg(\frac{1}{2}|\acute 0\rangle\langle u| +  \frac{1}{2}|\acute 0\rangle\langle d| + \frac{1}{\sqrt 2}|\acute 0\rangle\langle D| \bigg)e^{i \frac{\gamma^2_e}{2A} B^2_z t}e^{ - i\xi t} + {\rm H.c.}.
\end{eqnarray}

For the sake of simplicity in the presentation of the results, we will make the following assumptions: {\bf (i)} Since $\gamma_n\ll\gamma_e$ and $|\alpha|$, $|\beta|$, $|\gamma|$, $|\delta|$ are similar for the values of $B_z$ we  consider in the main text, the set of equations~(\ref{Scoefficients}) is
\begin{eqnarray}
c_{1 \acute{0}}=  \frac{1}{\gamma \beta - \alpha \delta}  (\gamma_e \gamma  + \gamma_n \delta) \approx \frac{\gamma_e\gamma}{\gamma \beta - \alpha \delta}  ,\nonumber\\
c_{1 0}=  \frac{-1}{\gamma \beta - \alpha \delta}  (\gamma_e \alpha  + \gamma_n \beta) \approx \frac{-\gamma_e\alpha}{\gamma \beta - \alpha \delta} ,\nonumber\\
c_{\acute{0} -1}=  \frac{-1}{\gamma \beta - \alpha \delta}  (\gamma_e \delta  + \gamma_n \gamma) \approx \frac{-\gamma_e \delta}{\gamma \beta - \alpha \delta},\nonumber\\
c_{0 -1}=  \frac{1}{\gamma \beta - \alpha \delta}  (\gamma_e \beta  + \gamma_n \alpha)\approx \frac{\gamma_e\beta}{\gamma \beta - \alpha \delta} .\nonumber\\
\end{eqnarray}
 And {\bf (ii)}, for the parameter regimes used in the main text one can further approximate 
\begin{eqnarray}
c_{1 \acute{0}}=  \frac{1}{\gamma \beta - \alpha \delta}  (\gamma_e \gamma  + \gamma_n \delta) \approx \frac{\gamma_e\gamma}{\gamma \beta - \alpha \delta} \approx  \frac{1}{\sqrt 2} \gamma_e,\nonumber\\
c_{1 0}=  \frac{-1}{\gamma \beta - \alpha \delta}  (\gamma_e \alpha  + \gamma_n \beta) \approx \frac{-\gamma_e\alpha}{\gamma \beta - \alpha \delta} \approx  \frac{-1}{\sqrt 2} \gamma_e,\nonumber\\
c_{\acute{0} -1}=  \frac{-1}{\gamma \beta - \alpha \delta}  (\gamma_e \delta  + \gamma_n \gamma) \approx \frac{-\gamma_e \delta}{\gamma \beta - \alpha \delta}  \approx  \frac{1}{\sqrt 2} \gamma_e,\nonumber\\
c_{0 -1}=  \frac{1}{\gamma \beta - \alpha \delta}  (\gamma_e \beta  + \gamma_n \alpha)\approx \frac{\gamma_e\beta}{\gamma \beta - \alpha \delta}\approx \frac{1}{\sqrt 2} \gamma_e .\nonumber\\
\end{eqnarray}
Hence, under {\bf (i)} and {\bf (ii)}, one can write the system Hamiltonian, i.e. Eq.~(\ref{Scleanbigreal}) as 
\begin{eqnarray}\label{cleanbigreal}
H &=& \omega_1 |1\rangle\langle 1| + \omega_{\acute 0} |\acute 0\rangle\langle \acute 0| + \omega_{-1} |-1\rangle\langle -1| + \omega_{0} | 0\rangle\langle 0| + \mu(t) \ \big(|1\rangle\langle 1| - |-1\rangle\langle -1|\big)\nonumber\\
&+&\sum_j \Omega_j  \bigg[  |1\rangle\langle \acute{0}| -  |1\rangle\langle 0| +  |\acute0\rangle\langle -1| +  |0\rangle\langle -1|  + {\rm H.c.} \bigg]   \cos{(\omega_j t +\phi_j)},
\end{eqnarray}
where $\Omega_j = \frac{B_x^j\gamma_e}{2\sqrt{2}}$. 
\subsection*{II A. Standard measurement scheme}\label{SSub1}
In order to remove magnetic field fluctuations from our sensor one can use two microwave fields resonant with the $0\leftrightarrow1$ and $0\leftrightarrow-1$ hyperfine transitions of the $^{171}$Yb$^{+}$  ion. Note that this is the scheme used in Refs.~\cite{Timoney11SM, Mikelsons15SM}. In particular, if one sets $\Omega=\Omega_{1}=\Omega_2$, $\phi_{1}=\pi$ and $\phi_2=0$, the following Hamiltonian  is obtained 
 (in the rotating frame of $H_0=\omega_1 |1\rangle\langle 1| + \omega_{\acute 0} |\acute 0\rangle\langle \acute 0| + \omega_{-1} |-1\rangle\langle -1| + \omega_{0} | 0\rangle\langle 0|$)
\begin{equation}\label{Sfirstnoisy}
H^I =  \mu(t) (|1\rangle\langle 1| - |-1\rangle\langle -1|) + \frac{\Omega}{2} (|1\rangle\langle 0| + |-1\rangle\langle 0| + {\rm H.c.}).
\end{equation}
In order to find the previous equation, one has to neglect terms rotating at a frequency $\propto \gamma_e B_z$ by invoking the  rotating wave approximation (RWA). 

The next step is to demonstrate how the addition of the two MW drivings leads to the cancellation of $\mu(t)$. For that, it is convenient to define a new basis $\{ |u\rangle, |d\rangle, |D\rangle, |\acute{0}\rangle \}$ such that 
\begin{eqnarray}\label{newbasis}
|u\rangle &=& \frac{1}{\sqrt{2}}(|B\rangle + |0\rangle),\nonumber\\
|d\rangle &=& \frac{1}{\sqrt{2}}(|B\rangle - |0\rangle),\nonumber\\
|D\rangle &=& \frac{1}{\sqrt{2}}(|-1\rangle - |1\rangle),
\end{eqnarray}
and $|B\rangle =\frac{1}{\sqrt{2}}(|1\rangle + |-1\rangle)$. In this new basis, Eq.~(\ref{Sfirstnoisy}) becomes
\begin{equation}
H =-\frac{\mu(t)}{\sqrt{2}} (|D\rangle\langle u| + |D\rangle\langle d|+ {\rm H.c.}) + \frac{\Omega}{\sqrt{2}} (|u\rangle\langle u| - |d\rangle\langle d|).
\end{equation}
Now, it is easy to see that, in the rotating frame of $\frac{\Omega}{\sqrt{2}} (|u\rangle\langle u| - |d\rangle\langle d|)$, the noisy term rotates at a speed $\propto \Omega$ and thus it can be eliminated with a suitable $\Omega$.

Let us now consider an additional rf-field signal interacting with the sensor, i.e we add an extra driving to Eq.~\eqref{cleanbigreal} whose Rabi frequency $(\Omega_{\rm tg})$ we want to determine. To this end, we use the energy difference between the $|\acute0\rangle \leftrightarrow |1 \rangle$ and $|\acute0\rangle \leftrightarrow |-1 \rangle$ transitions.  This energy difference is caused by the second-order Zeeman shift which is $\propto \frac{\gamma_e^2}{4A}B_z^2$ and, ideally, it would allow us to only excite the $|\acute0\rangle \leftrightarrow |1 \rangle$ transition. In this case, the general Hamiltonian is 
\begin{eqnarray}\label{rftarget}
H &=& \omega_1 |1\rangle\langle 1| + \omega_{\acute 0} |\acute 0\rangle\langle \acute 0| + \omega_{-1} |-1\rangle\langle -1| + \omega_{0} | 0\rangle\langle 0| + \mu(t) (|1\rangle\langle 1| - |-1\rangle\langle -1|)\nonumber\\
&+&\sum_j \Omega_j  \bigg[ |1\rangle\langle \acute{0}| - |1\rangle\langle 0| +  |\acute0\rangle\langle -1| +  |0\rangle\langle -1|  + {\rm H.c.} \bigg]   \cos{(\omega_j t +\phi_j)}\nonumber\\
&+&\Omega_{\rm tg} \  \bigg[  |1\rangle\langle \acute{0}| - |1\rangle\langle 0| +  |\acute0\rangle\langle -1| + |0\rangle\langle -1|  + {\rm H.c.} \bigg]   \cos{(\omega_{\rm tg} t +\phi_{\rm tg})}.
\end{eqnarray} 

In the rotating frame of $H_0=\omega_1 |1\rangle\langle 1| + \omega_{\acute 0} |\acute 0\rangle\langle \acute 0| + \omega_{-1} |-1\rangle\langle -1| + \omega_{0} | 0\rangle\langle 0|$  and selecting again $\Omega=\Omega_{1}=\Omega_2$, $\phi_{1}=\pi$ and $\phi_2=0$ one can find that the previous Hamiltonian becomes
\begin{eqnarray}\label{target}
H &=&-\frac{\mu(t)}{\sqrt{2}} (|D\rangle\langle u| + |D\rangle\langle d|+ {\rm H.c.}) + \frac{{\Omega}}{\sqrt{2}} (|u\rangle\langle u| - |d\rangle\langle d|)\nonumber\\
&+& \Omega_{\rm tg}  \bigg[  |1\rangle\langle \acute{0}| e^{i(\omega_1 - \omega_{\acute{0}})t} -  |1\rangle\langle 0| e^{i(\omega_1 - \omega_{0})t}+  |\acute0\rangle\langle -1| e^{i(\omega_{\acute0} - \omega_{-1})t} +  |0\rangle\langle -1| e^{i(\omega_{0} - \omega_{-1})t} + {\rm H.c.} \bigg]   \cos{(\omega_{\rm tg} t +\phi_{\rm tg})}.
\end{eqnarray}
If we tune $\omega_{\rm tg} = \omega_1 - \omega_{\acute0}$ and $\phi_{\rm rf} = 0$ in Eq.~(\ref{target}), and assuming that oscillating terms can be eliminated by the RWA, we would find  (note we have selected $\phi_{\rm tg} = 0$, but similar result can be derived for an arbitrary value of $\phi_{\rm tg}$)
\begin{eqnarray}\label{idealcase}
H &=&-\frac{\mu(t)}{\sqrt{2}} (|D\rangle\langle u| + |D\rangle\langle d|+ {\rm H.c.}) + \frac{\Omega}{\sqrt{2}} (|u\rangle\langle u| - |d\rangle\langle d|)+\frac{\Omega_{\rm tg}}{4} (|u\rangle\langle\acute0| + |d\rangle\langle\acute0| + {\rm H.c.})-\frac{\Omega_{\rm tg}}{2\sqrt{2}} (|D\rangle\langle\acute0| + |\acute0\rangle\langle D|),
\end{eqnarray}
which, in the rotating frame of $\frac{\Omega}{\sqrt{2}} (|u\rangle\langle u| - |d\rangle\langle d|)$, it adopts the following form
\begin{equation}\label{SHRabi}
H =-\frac{\Omega_{\rm tg}}{2\sqrt{2}} (|D\rangle\langle\acute0| + |\acute0\rangle\langle D|).
\end{equation}
Hamiltonian~(\ref{SHRabi}) corresponds to the Eq.~(2) given in the main text, from where it follows Eq. (3).  Recall that this is the approach followed in Ref.~\cite{Baumgart16SM}. 
\subsection*{II B. Refined measurement scheme}\label{SSub2}

The previous scheme  assumes several approximations that rely on the energy difference among the $|\pm 1\rangle$ and  $|\acute0\rangle$ states, and among  $|\pm 1\rangle$ and  $|0\rangle$. These energy differences are established by an external magnetic field, which also sets the frequency of the target rf-field that can be sensed. This is, when using the $|\acute0\rangle \leftrightarrow |1\rangle$ transition we can sense external fields of a frequency $\omega_{1} - \omega_{\acute0} \approx \frac{\gamma_e B_z}{2} - \frac{\gamma_e^2}{4A}B_z^2$ while, if we use the $|\acute0\rangle \leftrightarrow |-1\rangle$ spin transition, the $^{171}$Yb$^{+}$ sensor captures rf-radiation at a frequency $\omega_{\acute{0}} -  \omega_{-1} \approx \frac{\gamma_e B_z}{2} + \frac{\gamma_e^2}{4A}B_z^2$, see Eqs.~(\ref{Stransitions}). 

Both frequency differences  depend on the $B_z$ field magnitude. For example, in Ref.~\cite{Baumgart16SM}, $B_z$ is of the order of $\approx 1$ mT allowing to measure rf signals around $14$ MHz. Sensing signals with lower frequencies would require a reduction of the external magnetic field $B_z$ since $\omega_{1} - \omega_{\acute0}$ and $\omega_{\acute{0}} -  \omega_{-1}$ are proportional to $B_z$. However,  low values for $B_z$ leads to a weaker application of the RWA to the oscillating terms in Eq.~(\ref{rftarget}), thus to a failure of the whole sensing scheme. 

A more realistic approach should consider the following Hamiltonian 
\begin{eqnarray}
H &=& \omega_1 |1\rangle\langle 1| + \omega_{\acute 0} |\acute 0\rangle\langle \acute 0| + \omega_{-1} |-1\rangle\langle -1| + \omega_{0} | 0\rangle\langle 0| + \mu(t) ( |1\rangle\langle 1| -  |-1\rangle\langle -1|)\nonumber\\
&+& \Omega \bigg[  |1\rangle\langle \acute{0}| - |1\rangle\langle 0| +  |\acute0\rangle\langle -1| +  |0\rangle\langle -1|  + {\rm H.c.} \bigg]   \cos{(\omega^{\rm mw}_1 t + \phi_1)}\nonumber\\
&+& \Omega \bigg[  |1\rangle\langle \acute{0}| - |1\rangle\langle 0| +  |\acute0\rangle\langle -1| +  |0\rangle\langle -1|  + {\rm H.c.} \bigg]   \cos{(\omega^{\rm mw}_2 t + \phi_2)}\nonumber\\
&+& \Omega_{\rm tg} \bigg[  |1\rangle\langle \acute{0}| - |1\rangle\langle 0| +  |\acute0\rangle\langle -1| +  |0\rangle\langle -1|  + {\rm H.c.} \bigg]   \cos{(\omega_{\rm tg} t )}.
\end{eqnarray}
We proceed as in the previous subsection, that is, we move to a rotating frame w.r.t. the free-energy-terms $H_0= \omega_1 |1\rangle\langle 1| + \omega_{\acute 0} |\acute 0\rangle\langle \acute 0| + \omega_{-1} |-1\rangle\langle -1| + \omega_{0} | 0\rangle\langle 0|$. Furthermore, we select the MW control parameters such that 
$\omega_1^{\rm mw} = \omega_1 - \omega_0$, $\omega_2^{\rm mw} = \omega_{-1} - \omega_0$, $\phi_{1}=\pi$ and $\phi_2=0$. Then, if we neglect counter rotating terms oscillating at a GHz rate we have 
\begin{eqnarray}\label{theproblem}
H &=& \mu(t) ( |1\rangle\langle 1| -  |-1\rangle\langle -1|)+\frac{\Omega}{2} \bigg[  |1\rangle\langle 0| + |-1\rangle\langle 0| + {\rm H.c.} \bigg]  - \frac{\Omega}{2} \bigg[  |1\rangle\langle 0| e^{i\gamma_e B_z t} + |-1\rangle\langle 0| e^{-i\gamma_e B_z t}+ {\rm H.c.} \bigg] \nonumber\\
&+& \Omega_{\rm tg} \bigg[  |1\rangle\langle \acute{0}| e^{i(\omega_1 - \omega_{\acute 0})t} - |1\rangle\langle 0|  e^{i(\omega_1 - \omega_{0})t} +  |\acute0\rangle\langle -1| e^{i(\omega_{\acute{0}} - \omega_{-1})t} +  |0\rangle\langle -1|  e^{i(\omega_{0} - \omega_{-1})t}  + {\rm H.c.} \bigg]   \cos{(\omega_{\rm tg} t )}.
\end{eqnarray}
In the qubit basis~(\ref{newbasis}) the first line of the above Hamiltonian  transforms to  
\begin{eqnarray}
&&\frac{-\mu(t)}{\sqrt{2}} (|D\rangle\langle u| +  |D\rangle\langle d| + {\rm H.c.}) + \frac{\Omega}{\sqrt{2}}(|u\rangle\langle u| -  |d\rangle\langle d|)\nonumber\\
&-&   \bigg[\frac{\Omega}{2\sqrt{2}}(|u\rangle\langle u| -  |d\rangle\langle d|) +  \frac{\Omega}{4}  (|u\rangle\langle D| +  |D\rangle\langle d|) -  \frac{\Omega}{4}  (|D\rangle\langle u| +  |d\rangle\langle D|)  \bigg] e^{i\gamma_e B_z t} + {\rm H.c.}\nonumber\\
\end{eqnarray}

Now, if we want to use the $|\acute0\rangle \leftrightarrow |1\rangle$ spin transition as the detecting one and by taking into account that  there could be energy deviations $\xi$ in the frequency of the target signal of the kind $\omega_{\rm tg} = \omega_{1} - \omega_{\acute0} + \xi$, the second line of Hamiltonian~(\ref{theproblem}) is 
\begin{eqnarray}
&&\frac{\Omega_{\rm tg}}{2}( |1\rangle\langle \acute0| e^{-i\xi t} +  |\acute0\rangle\langle 1| e^{ i\xi t})\nonumber\\
&+&\frac{\Omega_{\rm tg}}{2} ( |1\rangle\langle \acute0| e^{2(\omega_1-\omega_{\acute0})t} e^{ i\xi t} + {\rm H.c.})\nonumber\\
&+&\frac{\Omega_{\rm tg}}{2} ( |\acute0\rangle\langle -1| e^{i (\omega_1 - \omega_{-1})t}e^{ i\xi t}  + {\rm H.c.})\nonumber\\
&+&\frac{\Omega_{\rm tg}}{2} ( |\acute0\rangle\langle -1|e^{i (2\omega_{\acute 0}-\omega_1 - \omega_{-1})t}e^{ - i\xi t} + {\rm H.c.}).
\end{eqnarray}
If we use the basis   $\{|u\rangle, |d\rangle, |D\rangle, |\acute{0}\rangle\}$ we get that the previous expression is 
\begin{eqnarray}
&&\bigg[\frac{\Omega_{\rm tg}}{4} (|u\rangle\langle\acute0| + |d\rangle\langle\acute0|) - \frac{\Omega_{\rm tg}}{2\sqrt{2}} |D\rangle\langle\acute0|\bigg]
e^{-i\xi t} + {\rm H.c.}\nonumber\\
&+&\frac{\Omega_{\rm tg}}{2} \bigg(\frac{1}{2} |u\rangle\langle \acute0| + \frac{1}{2} |d\rangle\langle \acute0| - \frac{1}{\sqrt{2}} |D\rangle\langle \acute0|  \bigg) e^{2i(\omega_1 - \omega_{\acute 0})t}e^{i\xi t} + {\rm H.c.}\nonumber\\
&+&\frac{\Omega_{\rm tg}}{2}\bigg(\frac{1}{2}|\acute 0\rangle\langle u| +  \frac{1}{2}|\acute 0\rangle\langle d| + \frac{1}{\sqrt 2}|\acute 0\rangle\langle D| \bigg)  e^{i (\omega_1 - \omega_{-1})t}e^{ i\xi t}  + {\rm H.c.}\nonumber\\
&+&\frac{\Omega_{\rm tg}}{2}\bigg(\frac{1}{2}|\acute 0\rangle\langle u| +  \frac{1}{2}|\acute 0\rangle\langle d| + \frac{1}{\sqrt 2}|\acute 0\rangle\langle D| \bigg)e^{i (2\omega_{\acute 0}-\omega_1 - \omega_{-1})t}e^{ - i\xi t} + {\rm H.c.}.
\end{eqnarray}

Then, the final target Hamiltonian is 
\begin{eqnarray}
H&=&\frac{-\mu(t)}{\sqrt{2}} (|D\rangle\langle u| +  |D\rangle\langle d| + {\rm H.c.}) + \frac{\Omega}{\sqrt{2}}(|u\rangle\langle u| -  |d\rangle\langle d|)\nonumber\\
&-&   \bigg[\frac{\Omega}{2\sqrt{2}}(|u\rangle\langle u| -  |d\rangle\langle d|) +  \frac{\Omega}{4}  (|u\rangle\langle D| +  |D\rangle\langle d|) -  \frac{\Omega}{4}  (|D\rangle\langle u| +  |d\rangle\langle D|)  \bigg] e^{i\gamma_e B_z t} + {\rm H.c.}\nonumber\\
&+&\bigg[\frac{\Omega_{\rm tg}}{4} (|u\rangle\langle\acute0| + |d\rangle\langle\acute0|) - \frac{\Omega_{\rm tg}}{2\sqrt{2}} |D\rangle\langle\acute0|\bigg]
e^{-i\xi t} + {\rm H.c.}\nonumber\\
&+&\frac{\Omega_{\rm tg}}{2} \bigg(\frac{1}{2} |u\rangle\langle \acute0| + \frac{1}{2} |d\rangle\langle \acute0| - \frac{1}{\sqrt{2}} |D\rangle\langle \acute0|  \bigg) e^{2i(\omega_1 - \omega_{\acute 0})t}e^{i\xi t} + {\rm H.c.}\nonumber\\
&+&\frac{\Omega_{\rm tg}}{2}\bigg(\frac{1}{2}|\acute 0\rangle\langle u| +  \frac{1}{2}|\acute 0\rangle\langle d| + \frac{1}{\sqrt 2}|\acute 0\rangle\langle D| \bigg)  e^{i (\omega_1 - \omega_{-1})t}e^{ i\xi t}  + {\rm H.c.}\nonumber\\
&+&\frac{\Omega_{\rm tg}}{2}\bigg(\frac{1}{2}|\acute 0\rangle\langle u| +  \frac{1}{2}|\acute 0\rangle\langle d| + \frac{1}{\sqrt 2}|\acute 0\rangle\langle D| \bigg)e^{i (2\omega_{\acute 0}-\omega_1 - \omega_{-1})t}e^{ - i\xi t} + {\rm H.c.}.
\end{eqnarray}
Or, if we use the relations (cf. {\bf Section~I})
\begin{eqnarray}
\omega_{1} - \omega_{\acute0} &\approx& \frac{\gamma_e B_z}{2} - \frac{\gamma_e^2}{4A}B_z^2,\nonumber\\
\omega_{1} - \omega_{0}&\approx& A + \frac{\gamma_e B_z}{2} + \frac{\gamma_e^2}{4A}B_z^2,\nonumber\\
\omega_{\acute{0}} -  \omega_{-1}&\approx& \frac{\gamma_e B_z}{2} + \frac{\gamma_e^2}{4A}B_z^2,\nonumber\\
\omega_{-1} -  \omega_{0}&\approx& A - \frac{\gamma_e B_z}{2} + \frac{\gamma_e^2}{4A}B_z^2,
\end{eqnarray}
the above Hamiltonian reads
\begin{eqnarray}\label{eq:allH}
H&=&\frac{-\mu(t)}{\sqrt{2}} (|D\rangle\langle u| +  |D\rangle\langle d| + {\rm H.c.}) + \frac{\Omega}{\sqrt{2}}(|u\rangle\langle u| -  |d\rangle\langle d|)\nonumber\\
&-&   \bigg[\frac{\Omega}{2\sqrt{2}}(|u\rangle\langle u| -  |d\rangle\langle d|) +  \frac{\Omega}{4}  (|u\rangle\langle D| +  |D\rangle\langle d|) -  \frac{\Omega}{4}  (|D\rangle\langle u| +  |d\rangle\langle D|)  \bigg] e^{i\gamma_e B_z t} + {\rm H.c.}\nonumber\\
&+&\bigg[\frac{\Omega_{\rm tg}}{4} (|u\rangle\langle\acute0| + |d\rangle\langle\acute0|) - \frac{\Omega_{\rm tg}}{2\sqrt{2}} |D\rangle\langle\acute0|\bigg]e^{-i\xi t} + {\rm H.c.}\nonumber\\
&+&\frac{\Omega_{\rm tg}}{2} \bigg(\frac{1}{2} |u\rangle\langle \acute0| + \frac{1}{2} |d\rangle\langle \acute0| - \frac{1}{\sqrt{2}} |D\rangle\langle \acute0|  \bigg) e^{2i(\frac{\gamma_e B_z}{2} - \frac{\gamma_e^2}{4A}B_z^2)t}e^{i\xi t} + {\rm H.c.}\nonumber\\
&+&\frac{\Omega_{\rm tg}}{2}\bigg(\frac{1}{2}|\acute 0\rangle\langle u| +  \frac{1}{2}|\acute 0\rangle\langle d| + \frac{1}{\sqrt 2}|\acute 0\rangle\langle D| \bigg)  e^{i \gamma_e B_z t}e^{ i\xi t}  + {\rm H.c.}\nonumber\\
&+&\frac{\Omega_{\rm tg}}{2}\bigg(\frac{1}{2}|\acute 0\rangle\langle u| +  \frac{1}{2}|\acute 0\rangle\langle d| + \frac{1}{\sqrt 2}|\acute 0\rangle\langle D| \bigg)e^{i \frac{\gamma^2_e}{2A} B^2_z t}e^{ - i\xi t} + {\rm H.c.},
\end{eqnarray}
which is denoted as $H_{\rm r}$ and given in Eq. (A1) of the main text.

\section*{III. Deviation from Rabi oscillations}
From Eq.~\eqref{eq:allH}, one can already notice that the sensor will soon depart from displaying the ideal coherent Rabi oscillations predicted by Eq.~\eqref{SHRabi} when the rf-signal has either a low frequency such that the RWA cannot be safely applied, a possible detuning w.r.t. the resonant condition $\omega_{\rm tg}=\omega_1-\omega_{\acute 0}+\xi$ with $|\xi|\ll \omega_{\rm tg}$ s.t. $|\xi|>0$, and/or a large Rabi frequency, $\Omega_{\rm tg}\sim \Omega$.

\begin{figure}
  \includegraphics[width=0.95\columnwidth]{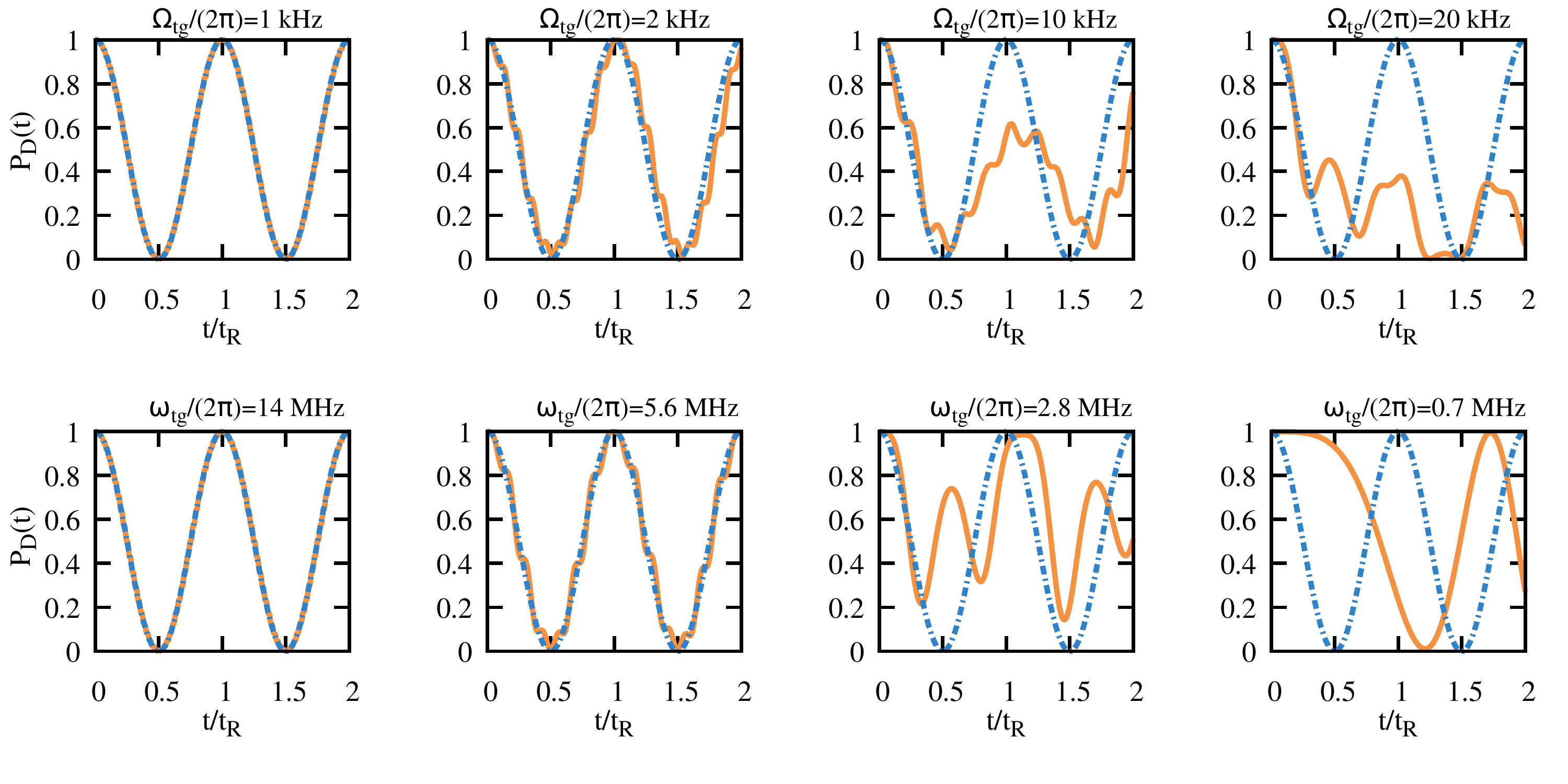}
  \caption{Dynamics of $P_{\rm D}(t)$ with $P_{\rm D}(0)=1$ using the full Hamiltonian $H_{\rm r}$, Eq.~(\ref{eq:allH}) (solid orange line) and its approximated expression, $P_{\rm D}(t)\approx \cos^2(\Omega_{\rm tg}t/\sqrt{8})$ (dashed blue line), for different parameter regimes and $\xi=0$. The left panels, top and bottom, correspond to $\omega_{\rm tg}\approx 2\pi\times 14$ MHz with $\Omega_{\rm tg}=2\pi\times 1$ kHz, a case within the validity discussed in~\cite{Baumgart16SM}. Top row shows the effect of increasing $\Omega_{\rm tg}$ (from left to right), i.e., $2$, $10$ and $20$ kHz (keeping fixed $\omega_{\rm tg}\approx 2\pi\times 14$ MHz). The bottom row shows the effect of reducing $B_z$ (i.e., reducing $\omega_{\rm tg}$ and keeping fixed $\Omega_{\rm tg}=2\pi \times 1$ kHz), from left to right: $5.6$, $2.8$ and $0.7$ MHz. \label{figparam} } 
\end{figure}

In particular, as discussed in~\cite{Baumgart16SM}, fields with a frequency of $\omega_{\rm tg}=2\pi\times 14$ MHz can be measured with high precision, whose amplitude can be up to few kHz, i.e. $\Omega_{\rm tg}\lesssim 2\pi \times 3.3$ kHz. Note that the Rabi frequencies for the microwave driving the transitions $\ket{0} \leftrightarrow \ket{1}$ and $\ket{0} \leftrightarrow \ket{-1}$ amount to $\Omega=2\pi\times 37.27$ kHz, which grants a robust decoupling w.r.t. magnetic field fluctuations. In Fig.~\ref{figparam} we show the evolution of $P_{\rm D}(t)$ for different parameters s.t. $P_{\rm D}(0)=1$, keeping $\xi=0$ and starting with $\Omega_{\rm tg}=2\pi\times 1$ kHz and $B_z=10$ G ($\omega_{\rm tg}\approx 2\pi\times 14$ MHz) in which $P_{\rm D}(t)$ can be well approximated by $P_{\rm D}(t)\approx \cos^2(\Omega_{\rm tg}t/\sqrt{8})$, as used in~\cite{Baumgart16SM} which follows from Eq.~\eqref{SHRabi}, and as function of time rescaled by $t_R=2\sqrt{2}\pi/\Omega_{\rm tg}$ (the time of a full Rabi oscillation within the approximated dynamics). Then, either increasing $\Omega_{\rm tg}$ (top panels) or decreasing $B_z$ (i.e. $\omega_{\rm tg}$) (bottom panels) leads to a departure from the RWA and more structured dynamics are observed. See caption for the considered parameters.

\begin{figure}
  \includegraphics[width=0.7\columnwidth]{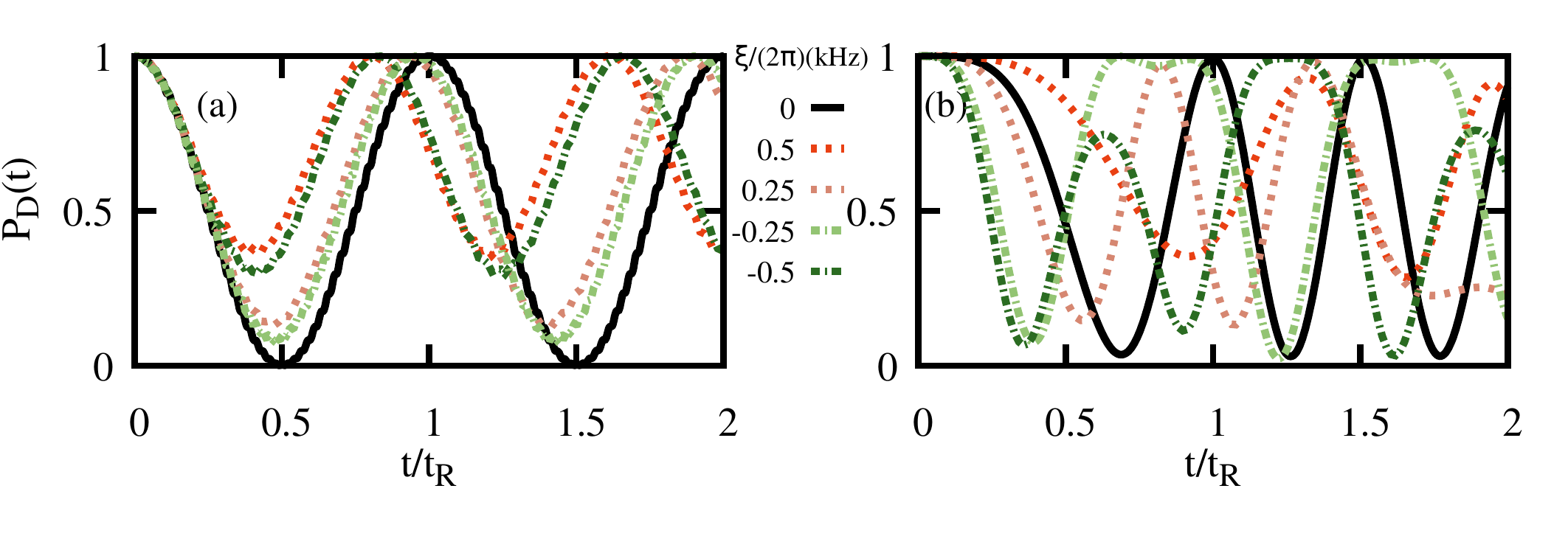}
  \caption{ Dynamics of $P_{\rm D}(t)$ with $P_{\rm D}(0)=1$ using the full Hamiltonian $H_{\rm r}$, Eq.~(\ref{eq:allH}), for (a) $\Omega_{\rm tg}=2\pi\times 1$ kHz and $\omega_{\rm tg}\approx 2\pi \times 14$ MHz and (b) $\Omega_{\rm tg}=2\pi\times 2$ kHz and $\omega_{\rm tg}\approx 2\pi \times 2.8$ MHz, for different detunings $\xi/(2\pi)=0$ (solid black), $0.5$ kHz (dotted red), $0.25$ kHz (dotted light red), $-0.25$ kHz (dashed light green) and $-0.5$ kHz (dashed dark green).\label{figdet} } 
\end{figure}

The impact of a detuned signal with respect to the resonant frequency splitting $\omega_1-\omega_{\acute 0}$ by an amount $\xi$ is illustrated in  Fig.~\ref{figdet}. We show two cases, namely, when the RWAs can be safely applied ($\Omega_{\rm tg}=2\pi\times 1$ kHz and $\omega_{\rm tg}\approx 2\pi \times 14$ MHz) (cf. Fig.~\ref{figparam}) and for a case in which the dynamics is more structured ($\Omega_{\rm tg}=2\pi\times 2$ kHz and $\omega_{\rm tg}\approx 2\pi \times 2.8$ MHz). For larger detunings, the rf-signal is not capable of producing transitions in the sensor, and thus the population remains constant $P_D(t)\approx 1$.


\begin{figure}
  \includegraphics[width=0.95\columnwidth]{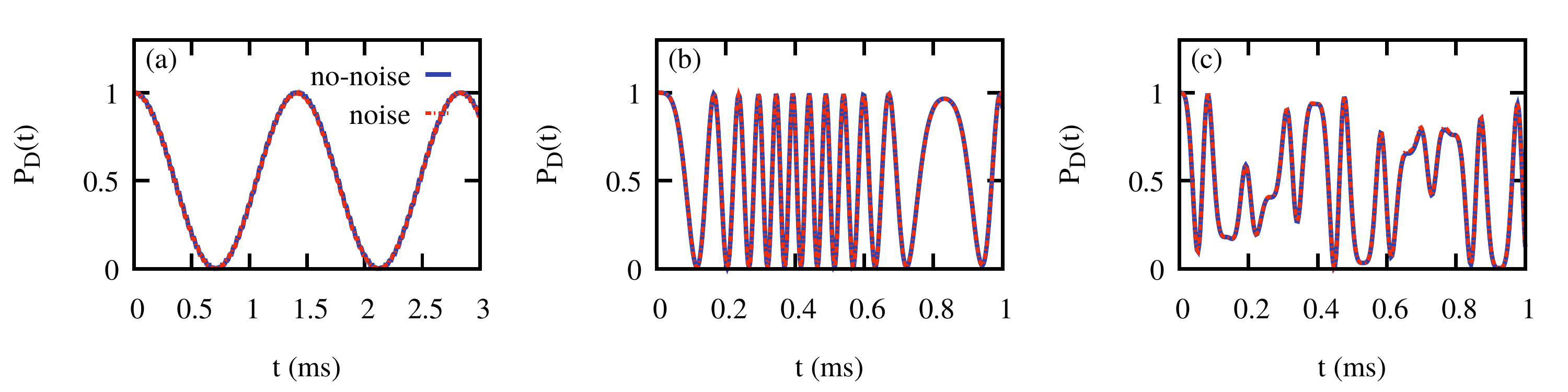}
  \caption{Dynamics of $P_{\rm D}(t)$ with $P_{\rm D}(0)=1$ using the full Hamiltonian $H_{\rm r}$, Eq.~(\ref{eq:allH}) with (dashed red line) and without (solid blue line) noises (magnetic-field and intensity fluctuations), as commented in {\bf Section~IV}. The results are equivalent, i.e., these noise sources do produce a significant impact in this time scale. The noisy dynamics has been obtained averaging $100$ stochastic repetitions. Panel (a) corresponds to the case I in main text, for $\Omega_{\rm tg}=2\pi\times 1$ kHz, $\omega_{\rm tg}=2\pi\times 14$ MHz and $\xi=0$, while (b) and (c) clearly deviate from the coherent Rabi oscillations, with $\Omega_{\rm tg}=2\pi\times 15$ kHz, $\omega_{\rm tg}=2\pi\times 2.8$ MHz and $\xi=0$, and $\Omega_{\rm tg}=2\pi\times 12$ kHz, $\omega_{\rm tg}=2\pi \times 7$ MHz and $\xi=2\pi\times 0.1 $ kHz (cf. Fig. 3 in the main text), respectively.  \label{figfluct} } 
\end{figure}
\section*{IV. Magnetic-field and amplitude fluctuations}\label{s:fluct}
The quantum sensor is prone to magnetic-field as well as intensity fluctuations of the Rabi frequencies. The magnetic-field fluctuations enter in the Hamiltonian as $\mu(t)(|1\rangle \langle 1|-|-1\rangle \langle -1|)$, that transforms in the final Hamiltonian to terms producing spurious transitions in the subspace spanned by $\{|u\rangle,|d\rangle,|D\rangle\}$ (cf. Eq. (A1) in the main text or Eq.~\eqref{eq:allH} here). Such fluctuations can be well described by a stochastic Orstein-Uhlenbeck process $\mu(t)$~\cite{Uhlenbeck30SM,Gillespie96aSM,Gillespie96bSM}. This Gaussian noise is fully characterized by its correlation time $\tau_\mu$ and intensity $\sigma_\mu$, with $\langle \mu(t)\rangle=0$ and $\langle\mu(t+\delta t)\mu(t) \rangle=\sigma^2_\mu e^{-\delta t/\tau_\mu}$ for $\delta t\geq 0$, and it allows for an exact update formula~\cite{Gillespie96aSM,Gillespie96bSM},
\begin{align}
\mu(t+\delta t)=\mu(t)e^{-\delta t/\tau_\mu}+\sigma_\mu\left(1-e^{-2\delta t/\tau_\mu} \right)^{1/2}N(t),
  \end{align}
with $N(t)$ denoting a random variable drawn from a normal distribution, $\langle N(t) \rangle=0$ and $\langle N(t)N(t+t') \rangle=\delta(t-t')$. This noise fulfills the properties of a continuous Markov process. From the previous update formula, one can calculate
\begin{align}
\langle \zeta^2(t) \rangle=\sigma^2_\mu\tau_\mu^2\left[\frac{2t}{\tau_m}-3+4e^{-t/\tau_m}-e^{-2t/\tau_m} \right]
  \end{align}
with $\zeta(t)=\int_0^t ds \ \mu(s)$. In this manner, it is easy to see that a state prepared in a $|1\rangle \pm |-1\rangle$ superposition evolving under $\mu(t)(|1\rangle \langle 1|-|-1\rangle \langle -1|)$ will decay as $\langle \sigma_{x;1,-1}(t) \rangle=e^{-\frac{1}{2}\langle \zeta^2(t)\rangle}$, where $\sigma_{x;1,-1}=|1\rangle \langle -1|+{\rm H.c.}$.  The decoherence time induced by these magnetic-field fluctuations is defined as $\langle \sigma_{x;1,-1}(T_2) \rangle=e^{-1}$, so that
\begin{align}
\sigma_\mu=\left(\tau_\mu (T_2-\tau_\mu(3/2-2e^{-T_2/\tau_\mu}+1/2 e^{-2T_2/\tau_\mu}) \right)^{-1}.
  \end{align}
For an exponential decay of the coherence, as typically observed in experiments, $\langle \sigma_{x;1,-1}(t) \rangle \propto e^{-t/T_2}$, one obtains the condition $\tau_m\ll T_2$, which in turn leads to $\sigma\approx 1/(T_2\tau_m)$. Here we have used $T_2=5.3$ ms as measured in~\cite{Baumgart16SM}, and $\tau_\mu=T_2/100$. For the intensity field fluctuations we include $\Omega\rightarrow \Omega(1+\epsilon(t))$ where $\epsilon(t)$ again follows an Orstein-Uhlenbeck process with $\tau_\epsilon=1$ ms and relative intensity of $2.5\times 10^{-3}$, as given in~\cite{Cai12SM}. These two sources of noise do not produce a significant impact in the dynamics of the populations in the time scale considered here (see Ref.~\cite{Baumgart16SM} for experimental results). See Fig.~\ref{figfluct} for examples showing the dynamics of the population $P_D(t)$ with and without including these noise sources.

\section*{V. Simulation of an experimental acquisition}
Let us denote the population we are interested in measuring at time $t_k$ by $P_k$. When the quantum sensor is interrogated after an evolution time $t_k$, one retrieves $1$ with probability $P_k$ (when found in $\ket{D}$, as considered in the main text), and $0$ with probability $1-P_k$. This binomial process allows us to obtain an estimate of $P_k$ after repeating the measurement $N_m$ times, which we denote here by $P^s_k$ and reads as
\begin{align}
  P_k^s=\frac{1}{N_{m}}X_k=\frac{1}{N_{m}}\sum_{n=1}^{N_m}x_{n;k},
  \end{align}
where $x_{n;k}$ is the $n$th outcome, i.e., a random variable drawn from a Binomial distribution $B(1,P_k)$ with success probability $P_k$, such that $x_{n;k}=\{0,1\}$ and $X_k=\sum_{n=1}^{N_p}x_{n;k}$ the number of $1$'s recorded at the interrogation time $t_k$. Only for illustration purposes, we assign a shot-noise uncertainty to each $P_k^s$, which is given by $\sigma_k=\max(1/N_{m},\sigma_{N_{m};k}/\sqrt{N_m})$, with $\sigma_{n;k}$ the standard deviation of the list of outcomes $\{x_{1;k},x_{2;k},\ldots,x_{N_m;k}\}$. In the limit of many outcomes, $N_m\gg 1$, it will read as $\sigma_{N_m;k}=\sqrt{P_k(1-P_k)}$ so that $\sigma_{N_{m};k}/\sqrt{N_m}$ is the standard error of mean and $\sigma_k=\sigma_{N_m;k}/\sqrt{N_m}$. Note that the uncertainty $1/N_{m}$ gives account of the variation in the estimate $P_k^s$ if one value is flipped, $x_{n;k}\rightarrow 1-x_{n;k}$. More precisely, this uncertainty stems from the confidence interval in determining $P_k^s$ that after $N_m$ trials any has not been successful. With a $68.2\%$ confidence interval (equivalent to $1\sigma$ in a normal distribution)  the probability of $P_k$ being $0$ reads as $p(P_k^s=0)\leq 0.318$, which for $P_k$ close to $0$, it follows that $0\leq P_k\leq -\log(0.318)/N_m$, which can be approximated to $0\leq P_k\leq 1/N_m$. In a similar manner, if all of the $N_m$ trials have been successful, the same argument applies.

For reproducibility, we provide in the following the string of outcomes obtained randomly and used for the analysis shown in the main text, for both cases. For {\em Case I}, with $t_k=2.83 (k-1)/(N_p-1)$ ms for $k=1,\ldots,N_p$ and $N_p=18$, we use $X_k=\{1,1,0,0,0,1,1,0,1,0,0,0,0,0,0,1,1,1\}$ for $N_m=1$,  $X_k=\{4,3,1,1,0,0,0,2,4,3,4,2,0,0,1,2,2,4\}$ for $N_m=4$, and $X_k=\{20,18,11,4,0,2,11,12,18,20,12,7,0,0,7,9,18,20\}$ for $N_m=20$. 
For {\em Case II}, with $t_k=0.236 (k-1)/(N_p-1)$ ms with $N_p=20$, we use $X_k=\{4,4,2,2,1,0,4,3,2,1,0,1,0,1,3,2,1,0,2,1\}$ for $N_m=4$,  $X_k=\{20,20,16,11,0,6,16,20,11,5,6,2,2,3,2,9,12,7,4,7\}$ for $N_m=20$, and $X_k=\{40,40,34,22,7,14,35,38,22,13,5,5,6,8,10,22,22,12,8,11\}$ for $N_m=40$. As commented in the main text, and presented below, we also consider a single-shot acquisition for this case, i.e. $X_k=\{1,1,1,1,0,1,1,1,1,0,0,0,0,0,0,1,0,1,1,0\}$ for $N_m=1$.

%, the quantum sensor can be found in that state with probability $p_k$, while with probability $1-p_k$ the system is not in that  particular state. 

\section*{VI. Bayesian Inference}
Here we provide more examples and details of the numerical calculations presented in the main text where Bayes' rule is used to provide reliable estimators of unknown parameters based on the observed/measured data~\cite{vonderLindenSM,GelmanSM}.

From Bayes' theorem we know that the posterior probability $p({\bf \Theta}|{\bf D})$  is proportional to  $p({\bf D}|{\bf \Theta})p({\bf \Theta})$ (up to a normalization factor). This probability distribution contains the information we can extract from the observed data ${\bf D}$ given the prior knowledge over the parameters $p({\bf \Theta})$, and the likelihood $p({\bf D}|{\bf \Theta})$. As commented in the main text,  the Binomial statistics of $X_k$ leads to
\begin{align}\label{sm:likelihood}
  p({\bf D}|{\bf \Theta})= \Pi_{k=1}^{N_p}f(X_k,N_m,\tilde{P}_k(t_k;{\bf \Theta}) \quad {\rm with} \quad f(x,n,p)=\frac{n!}{x!(n-x)!}p^x(1-p)^{(n-x)},
%=\Pi_{k=1}^{N_p}\mathcal{N}(P^s_k(t_k)-\tilde{P}_k(t_k;{\bf \Theta}),\sigma_k^2),
\end{align}
where  $f(x,n,p)$ denotes a the probability of having observed $x$ success outcomes from $n$ trials from a Binomial distribution with success probability $p$, while $\tilde{P}_k(t_k;{\bf \Theta})$ stands for the expected population at time $t_k$ when using ${\bf \Theta}$ as the parameters in the model. The values $X_k$ from $N_m$ measurements at each time $t_k$ form the observations, i.e., the data ${\bf D}$. In the following we provide more details about the first case of study presented in the main text, namely, {\em case I}, while more information on Markov Chain Monte Carlo methods, relevant for the {\em case II}, is presented in {\bf Section~VII}.

\subsection*{VI A. Case I}
\begin{figure}
\centering
\includegraphics[width=1\linewidth,angle=-0]{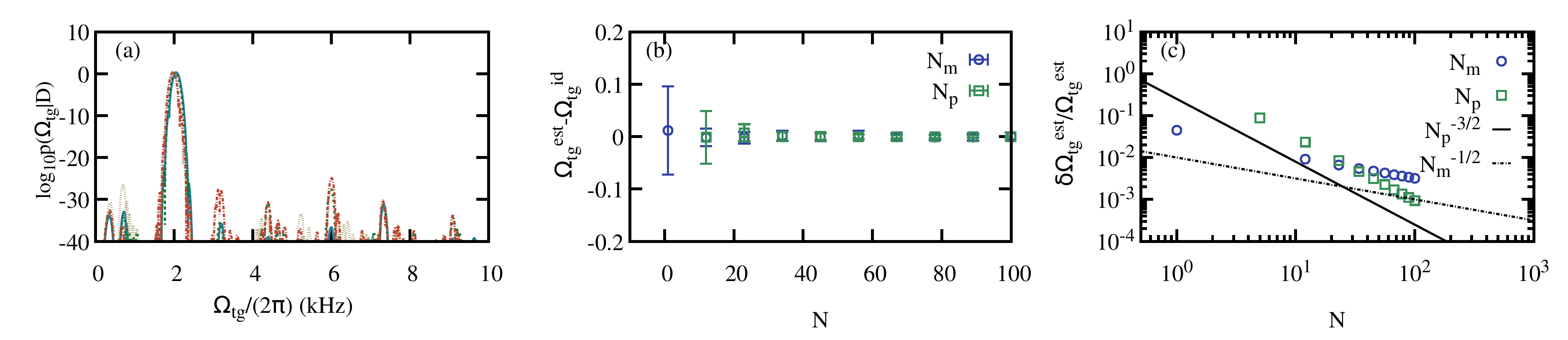}
\caption{\small{(a) Logarithm of the posterior probability distribution over the amplitude $\Omega_{\rm tg}$, $\log_{10}p(\Omega_{\rm tg}|{\bf D})$ given four different and independent observations ${\bf D}$, with $N_p=21$, $N_m=5$ and generated from $\omega_{\rm tg}=2\pi\times 14$ MHz, $\xi=0$ and $\Omega_{\rm tg}^{\rm id}=2\pi\times 2$ kHz. (b) Difference between $\Omega_{\rm tg}^{\rm est}$, obtained upon an average of $40$ runs (i.e. over $40$ independent sets of observations ${\bf D}$), and $\Omega_{\rm tg}^{\rm id}$ (in kHz), as function of $N=\{N_m,N_p\}$, namely, increasing the number of measurements $N_m$ per point and keeping $N_p=21$ points and $t_f=3/\sqrt{2}$ ms (blue circles), and increasing the number of points $N_p$ (and thus $t_f'=t_f N_p/21$ so that $t_{k+1}-t_k\approx 0.1$ ms) keeping $N_m=10$ measurements per point (green squares). In panel (c) we show the precision of the estimated Rabi frequency $\delta\Omega_{\rm tg}^{\rm est}/\Omega_{\rm tg}^{\rm est}$ as $N=\{N_m,N_p\}$ increases. The lines are guides to the eyes marking different scaling with $N$.}}
\label{sm:cossignal}
\end{figure}

The application of the RWAs allows us to obtain an analytical expression $P_D(t)=\cos^2(\Omega_{\rm tg}t/\sqrt{8})$ when the initial state is $\ket{D}$, i.e. $P_D(0)=1$~\cite{Baumgart16SM}. In this manner, one can easily compute Eq.~\eqref{sm:likelihood} and so the posterior by scanning different values of $\Omega_{\rm tg}$. Note that in this case, ${\bf \Theta}=\{ \Omega_{\rm tg}\}$ and  $\tilde{P}_k(t_k;{\bf \Theta})\rightarrow \cos^{2}(\Omega_{\rm tg}t_k/(2\sqrt{2}))$. This is plotted in Fig.~\ref{sm:cossignal}(a) for four different sets of observations ${\bf D}$ containing $N_p=21$ points, each of them obtained averaging $N_m=5$ measurements, and generated from $\omega_{\rm tg}=2\pi\times 14$ MHz, $\xi=0$ and $\Omega_{\rm tg}^{\rm id}=2\pi\times 2$ kHz. The time separation between consecutive points is $t_{k+1}-t_k\approx 0.1$ ms, so that we consider $\Omega_{\rm tg}$ up to $10$ kHz (see below for a discussion). Note that reliable estimates can be obtained even in the situation of reduced number of measurements and few recorded times (cf. Fig.~\ref{sm:cossignal}(b) and (c)). %In particular, for $N_p=10$ (so that $t_f\approx 1$ ms) and $N_m=10$, the $\Omega_{\rm tg}^{\rm est}=2\pi\times 2.002(44)$ kHz. 

\begin{figure}
\centering
\includegraphics[width=0.7\linewidth,angle=-0]{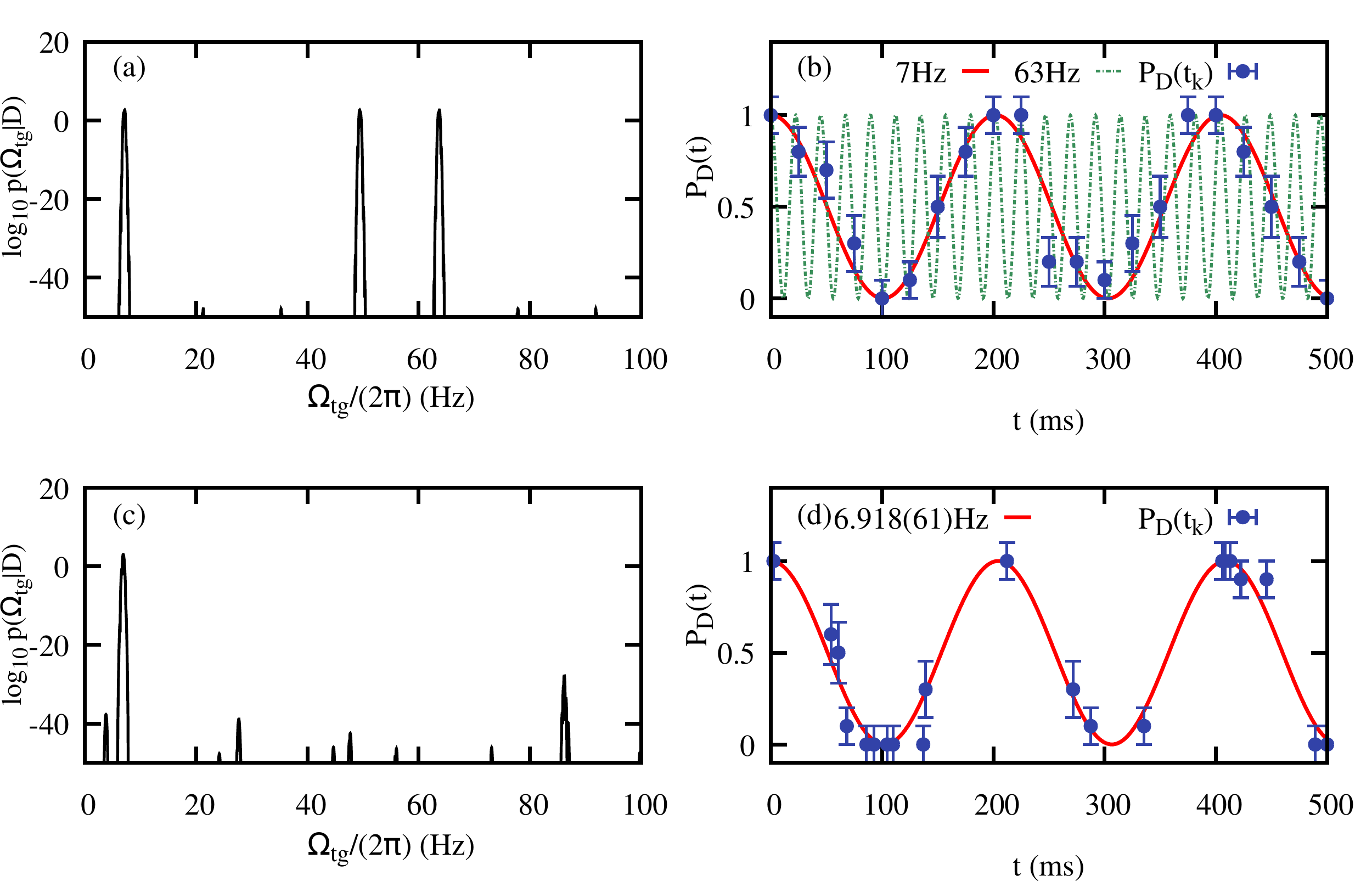}
\caption{\small{(a) Logarithm of the posterior probability distribution over the amplitude $\Omega_{\rm tg}$, $\log_{10}p(\Omega_{\rm tg}|{\bf D})$ given the data ${\bf D}$ in the panel (b), assuming an uninformative prior in $\Omega_{\rm tg}\in 2\pi[0.1,100]$ Hz. Note that the data has been obtained with $\Omega_{\rm tg}=2\pi\times 7$ Hz, so that the  peaks appearing at $2\pi\times 50$ and $2\pi\times 63$ Hz are spurious. In (b) we show the equally-space observed data (points) and the ideal signal $\Omega_{\rm tg}=2\pi \times 7$ Hz, and one of the resulting overfitted predictions, e.g. $2\pi \times 63$ Hz. Same plots in (c) and (d) but using a random time sampling in $t\in[0,t_f]$. In (d) the observed data ${\bf D}$, i.e. $P_k^s(t_k)$ with uncertainty $\sigma_k$, is plotted together with the predicted one $P_D(t)=\cos^2(\Omega_{\rm tg}^{\rm est}t/\sqrt{8})$ (red). See text for further details.}}
\label{sm:aliasing}
\end{figure}

% aliasing/random chosen times
It is well known that an equally-spaced sampling of a periodic signal can produce aliasing effects. In particular, if the time difference between two measured points is $\Delta t$, then the maximum frequency than can be inferred without aliasing is half of the sampling rate, set by the Nyquist frequency $f_f=1/(2\Delta t)$. For a periodic signal $y(t)\propto \cos(\omega t)$, this leads to $\omega^{\rm max}=2\pi/(2\Delta t)$. In our case, from Eq. (3) of the main text, one finds $\Omega_{\rm tg}^{\rm max}=2\pi/(\sqrt{2}\Delta t)$. For the {\em case I} shown in the main text, $\Delta t\approx 1/6$ ms, so that $\Omega_{\rm tg}^{\rm max}\approx 2\pi\times 6/\sqrt{2}$ kHz$\approx 2\pi \times 4.2$ kHz. This pre-knowledge can be included in the prior such that frequencies $\Omega_{\rm tg}>\Omega_{\rm tg}^{\rm max}$ are not considered, i.e. $p(\Omega_{\rm tg}>\Omega_{\rm tg}^{\rm max})=0$. See Fig.~\ref{sm:aliasing} for an illustration of this effect. The noisy signal has been obtained simulating an experiment with $N_p=21$ equally-spaced points, $N_{m}=10$ measurement repetitions per point and $\Omega_{\rm tg}=2\pi\times 7$ Hz, fixing $\xi=0$ and $\omega_{\rm tg}=2\pi\times 14$ MHz, from $t=0$ to $t_f=500$ ms, similar to one case explored in~\cite{Baumgart16SM}.  For an initial state $\ket{D}$, it follows $P_D(t)\approx \cos^2(\Omega_{\rm tg}t/\sqrt{8})$. Scanning the range $\Omega_{\rm tg}\in 2\pi [0.1,100]$ Hz, one finds that the posterior $p(\Omega_{\rm tg}|{\bf D})$ features three peaks at the values $2\pi\times 7$, $2\pi\times 50$ and $2\pi\times 63$ Hz. Taking into account these overfitted solutions, one finds $\Omega_{\rm tg}^{\rm est}\approx 2\pi \times 40(24)$ Hz. A simple post analysis however  allows us to discard such overfitted solutions or  frequencies above $\Omega_{\rm tg}^{\rm max}$ (cf. Fig.~\ref{sm:aliasing}(b)) by discarding values $\Omega_{\rm tg}>2\pi/(\sqrt{2}\Delta t)$, so that one obtains $\Omega_{\rm tg}^{\rm est}\approx 2\pi\times 7.147(42)$ Hz. We comment that the impact of such aliasing effects can be reduced through non-equal time sampling, such as randomly selecting the instances $t_k$ at which the sensor is interrogated. See Fig.~\ref{sm:aliasing} for the same parameters as before but where the sampling times $t_k$ have been selected randomly in the interval $[0,t_f]$. In this manner, with the data shown in Fig.~\ref{sm:aliasing}(d), one finds $\Omega_{\rm tg}^{\rm est}=2\pi \times 6.918(61)$ Hz when inspecting in $\Omega_{\rm tg}\in 2\pi [0.1,100]$ Hz.

It is worth mentioning that standard least-squares regression techniques can be applied here to fit the observations or data ${\bf D}$ to $P_{D}(t)=\cos^2(\Omega_{\rm tg} t/\sqrt{8})$. Note however that the uncertainties $\sigma_k$ will be in general different for each the $N_p$ points. Moreover, any pre-knowledge or bias about $\Omega_{\rm tg}$, i.e. any informative prior $p({\bf \Theta})$, makes the least-squares fit inapplicable. In addition, as posteriors need not be Gaussian, the Bayesian-inference based method allows us to gain more information of the unknown parameter.

\begin{figure}
\centering
\includegraphics[width=1\linewidth,angle=-0]{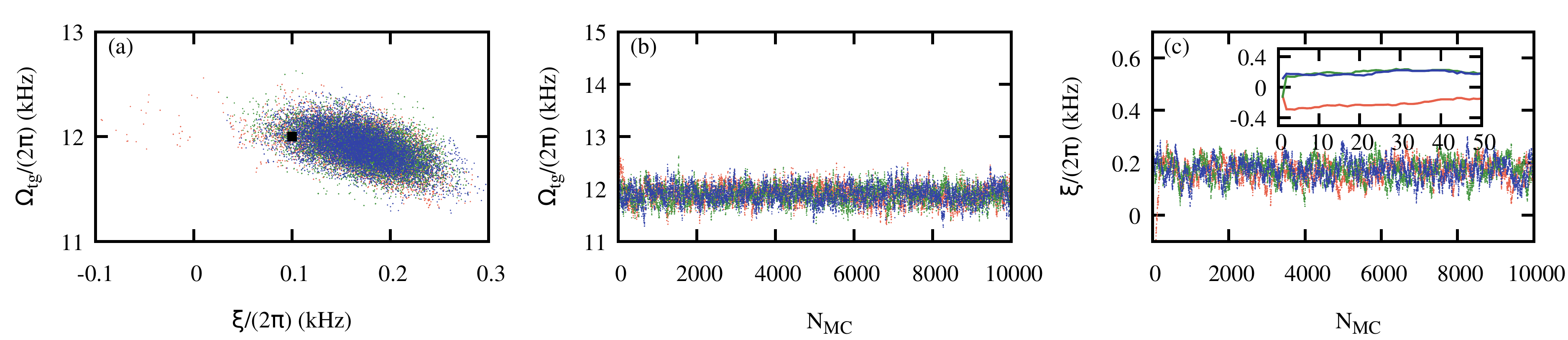}
\caption{\small{(a) Scatter plot of the recorded points during three independent MCMC as for the case considered in the main text (cf. Fig. (3)). Panels (b) and (c) illustrate the evolution of the chain as a function of the number of Monte Carlo steps. Inset in (c) show the burn-in regime for $\xi$, in which the initial points jump until reaching the convergence region with good mixing.}}
\label{sm:MCMC}
\end{figure}
%21.4325640102713  -0.273493421052428   7.69070653188804        0.461794919704818    2.26046788336339        -0.239165251854521

\section*{VII. Markov Chain Monte Carlo}\label{SSMCMC}
As explained in the main text, when the determination of the posterior probability distributions of the unknown parameters becomes complex and numerically demanding, one may resort to Markov Chain Monte Carlo (MCMC) methods~\cite{vonderLindenSM,GilksSM} for an efficient sampling of such distributions. Here we perform the sampling using a Metropolis algorithm, as explained in the main text. Fig.~\ref{sm:MCMC} shows an additional Markov chain for the same case considered in the Fig. (3) of the main text. The trace plots in Figs.~\ref{sm:MCMC}(b) and (c) show the evolution of the three independent Markov chains, which achieve a good convergence and mixing upon $100$ Monte Carlo steps. In order to speed up the convergence of the Markov chain, we perform $100$ Metropolis pre-steps  using the prior probability distributions to propose the subsequent step.  The results shown in the main text have been obtained removing the first $200$ steps of the MCMC after the $100$ of the (burn-in). The effective size of the Markov chain is approximately $N_{\rm MC}/2$. Recall that the prior probability distribution $p(\Omega_{\rm tg})$ is flat, i.e., uninformative in the region of interest, $0\leq \Omega_{\rm tg}\leq 2\pi\times 50$ kHz, while we take $p(\xi)=\mathcal{N}(0,\sigma_\xi^2)$ with $\sigma_\xi=2\pi\times 0.1$ kHz. The maximum value for the Rabi frequency is again related to the Nyquist frequency (see above) as $1/\Delta t\approx 80$ kHz for the {\em case II}, so that $\Omega_{\rm tg}^{\rm max}\approx 2\pi \times 57$ kHz. The steps during the MCMC are perform using $\sigma_\Omega^2$ and $\sigma_\xi^2$, where $\sigma_\Omega=2\pi\times 1$ kHz is found to give a good effective size. For slow converging cases one may rely to adaptive sampling, reducing both $\sigma_{\Omega}$ and $\sigma_\xi$, or by employing a different algorithm (e.g. Metropolis-Hastings)~\cite{GilksSM}.

\begin{figure}
\centering
\includegraphics[width=0.7\linewidth,angle=-0]{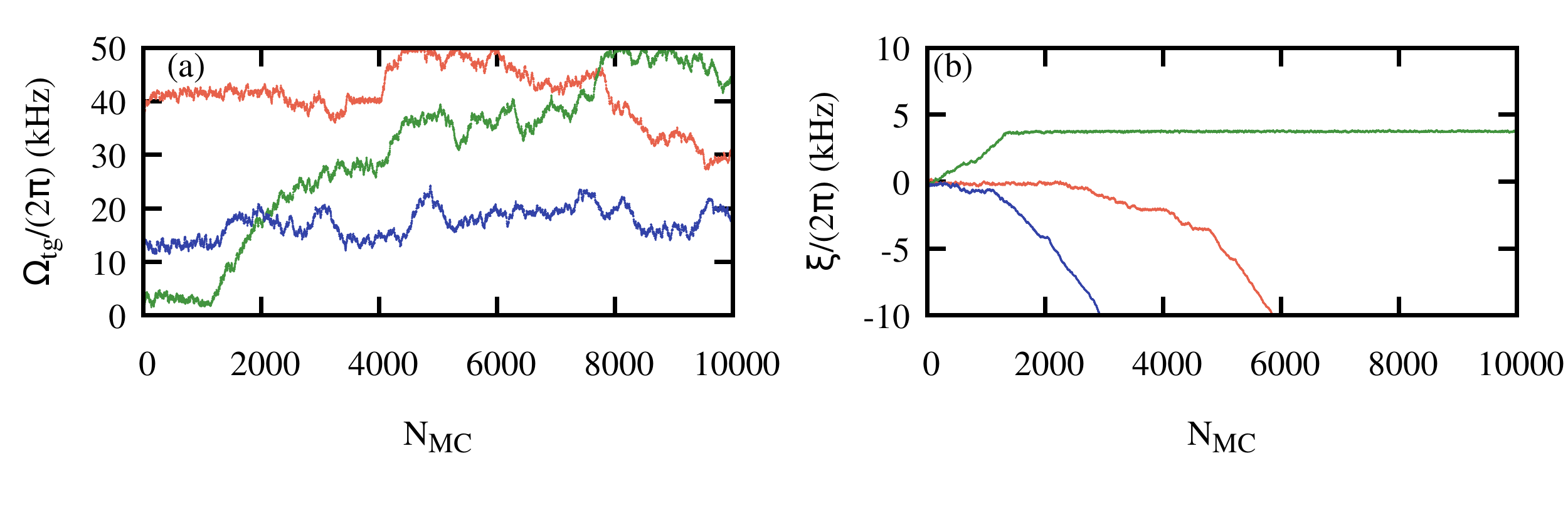}
\caption{\small{Results of three independent MCMC for a single shot acquisition ($N_m=1$) for the same parameters as used in Fig. 3(a) of the main text. The recorded MCMC steps for $\Omega_{\rm tg}$ (a) and $\xi$ (b). Due to the large shot noise, and the non-harmonic sensor response, the chains do not converge.}}
\label{figNm1caseII}
\end{figure}

Finally, we show in Fig.~\ref{figNm1caseII} the results of three independent MCMC when $N_m=1$ and same parameters as in Fig. 3 of the main text. Due to the large shot noise and the non-harmonic response of the quantum sensor, there is no convergence in the MCMC. Almost any pair of values ${\bf \Theta}=\{\Omega_{\rm tg},\xi\}$ provides a signal $\tilde{P}_k(t_k;{\bf \Theta})$ from which the observations ${\bf D}$ could have been obtained.

\begin{figure}
\centering
\includegraphics[width=0.7\linewidth,angle=-0]{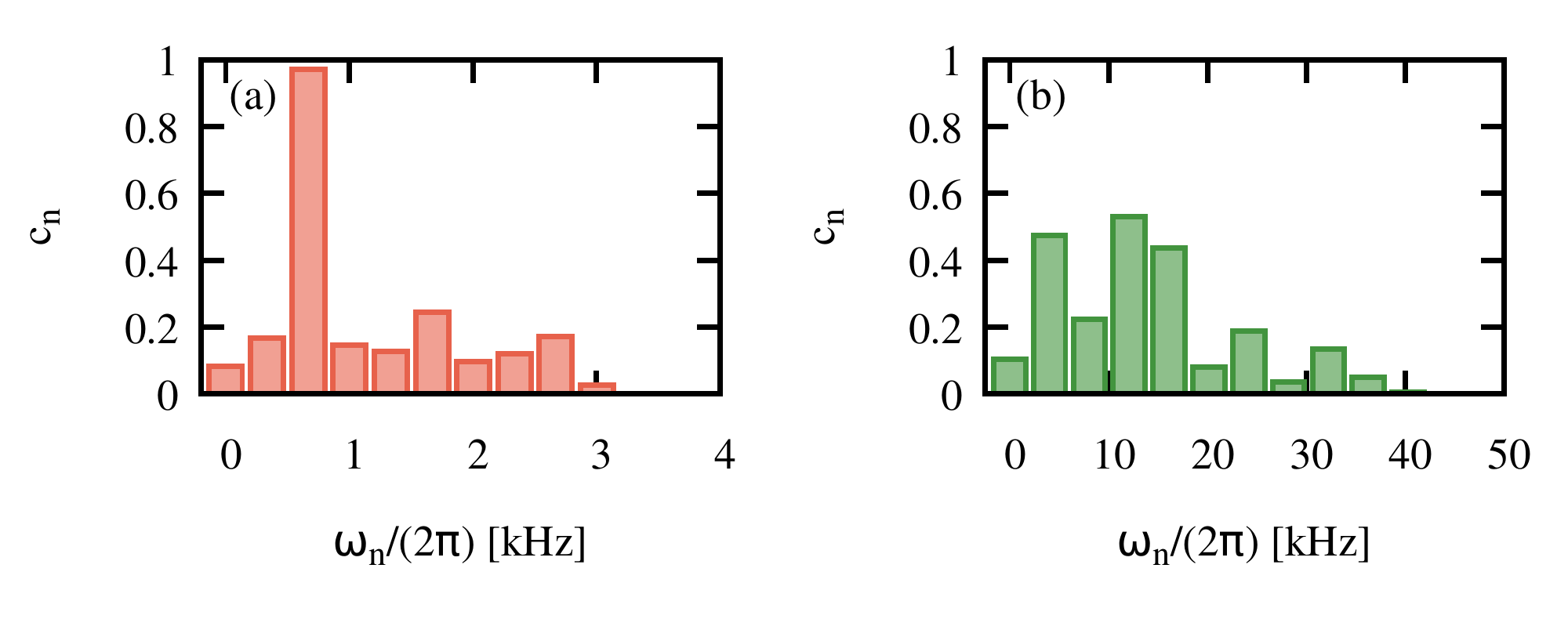}
\caption{\small{FFT spectrum of the data shown in Fig. 2(a) and Fig. 3(b) of the main text, plotted in (a) and (b), respectively. In (a), for the case I,  we identify a relevant frequency for which $c_n\approx 1$, whose value is $\omega_n=2\pi\times 0.6678$ kHz. In (b), which corresponds to the case II studied in the main text, the FFT provides different relevant frequencies, which together with the lack of a simple relation between $P_D(t)$ and $\Omega_{\rm tg}$ challenges the identification of $\Omega_{\rm tg}$, as well as of potential detunings $\xi$. }}
\label{figFFT}
\end{figure}
\section*{VIII. Fast Fourier Transform analysis and least-squares fits}
The Fast Fourier Transform (FFT) allows for the determination of the relevant frequencies of a signal. Here we show the results of the FFT for the two cases studied in the main text, namely, performing the FFT of the data ${\bf D}$ shown in Fig. 2(a) (case I) and Fig. 3(a) (case II). In particular, since the populations in ${\bf D}$ oscillate between $0$ and $1$, we shift and normalize the data to be withing $-1$ and $1$ to suppress the zero frequency component. In Fig.~\ref{figFFT} we show the spectrum of ${\bf D}$ on Fourier components $c_n$ at frequency $\omega_n$.

For the case I we find that $P_D(t)$ and $\Omega_{\rm tg}$ are related through Eq. (3) of the main text. Hence, one can obtain an estimate of $\Omega_{\rm tg}$ based on the FFT. In particular, here we see that the FFT of the data ${\bf D}$ leads to a predominant frequency with a weight close to one, thus revealing a monochromatic signal (cf. Fig.~\ref{figFFT}(a)). The maximum corresponds to $\omega_{\rm max}=2\pi\times 0.6678$ kHz. From $P_D(t)=\cos^2(\pi t/(2\pi\sqrt{2}/\Omega_{\rm tg}))$, it is easy to find $\Omega_{\rm tg}^{\rm est}=\sqrt{2}\omega_{\rm max}$. A rough uncertainty of this estimator is taken as $\delta \omega/4$ where $\delta \omega$ is the frequency resolution of the FFT, so that $\Omega_{\rm tg}^{\rm est}=2\pi\times 0.94(12)$ kHz. This estimated value, although compatible with the ideal one, $\Omega_{\rm tg}=2\pi\times 1$ kHz, is less accurate than the one obtained via Bayesian inference. In this case, a least-squares fit of the data ${\bf D}$ to the expression $P_D(t)=\cos^2(\Omega_{\rm tg}t/\sqrt{8})$ allows us to find estimates for $\Omega_{\rm tg}$. In particular, for $N_m=4$ measurements per point (see above for the actual string of outcomes)  we obtain $\Omega_{\rm tg}=2\pi\times 0.947(20)$. The uncertainty corresponds to a confidence interval of $68\%$, i.e. to $1\sigma$.

\begin{figure}
\centering
\includegraphics[width=0.7\linewidth,angle=-0]{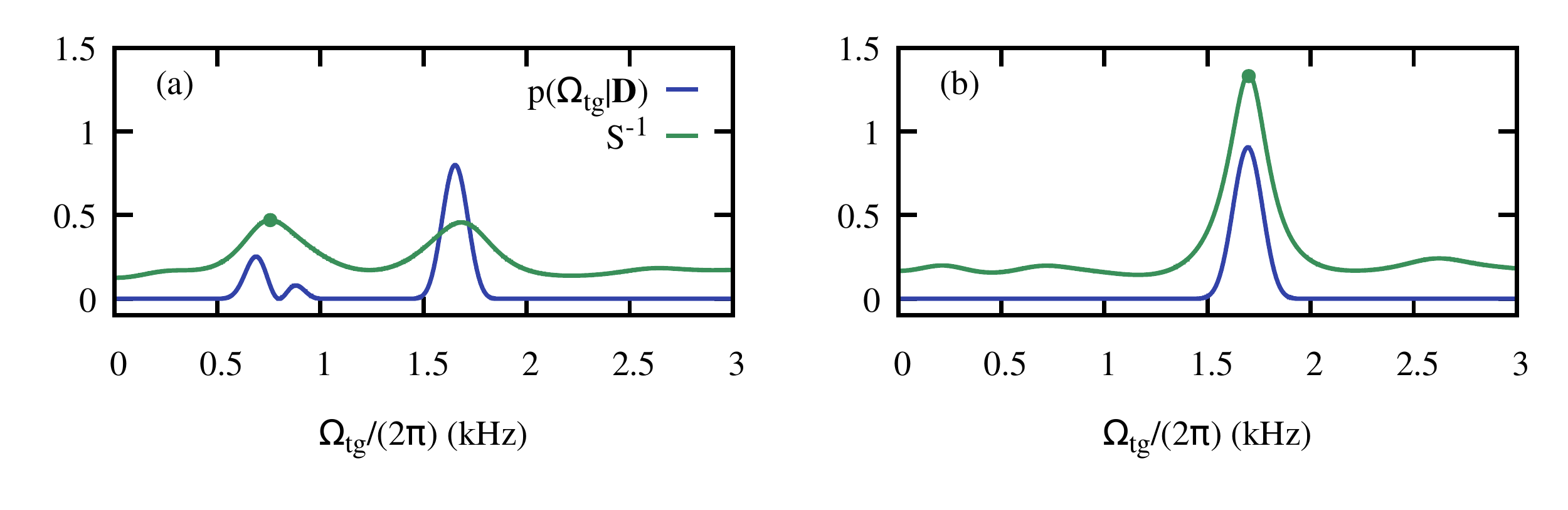}
\caption{\small{ Comparison between $S^{-1}$ for the least-squares fit (green) and the posterior distribution $p(\Omega_{\rm tg}|{\bf D})$ (blue) using a Bayesian analysis for two different realizations (a) and (b) with $\Omega_{\rm tg}=2\pi\times 1.6$ kHz and $N_m=1$ (single shot measurements) and $N_p=15$. The least squares maximizes $S^{-1}$. In both cases, the solid point indicates the estimator using least-squares $\Omega_{\rm tg}^{\rm est;LQ}$. In (a), the data ${\bf D}$ is such that $S^{-1}$ exhibits a maximum at $\Omega_{\rm tg}^{\rm est}\approx 2\pi \times 0.76$ kHz while the posterior still reveals a dominant contribution close to the true value $\Omega_{\rm tg}=2\pi\times 1.6$ kHz. In (b) we show a particular case in which the least-squares fit to the data ${\bf D}$ provides a more accurate estimator, $\Omega_{\rm tg}^{\rm est;LQ}=2\pi\times 1.699(57)$ kHz, than its Bayesian counterpart, $\Omega_{\rm tg}^{\rm est;Bayes}=2\pi\times 1.695(69)$ kHz. In average, however, we find $\overline{\delta \Omega_{\rm tg}^{\rm est;Bayes}/\delta \Omega_{\rm tg}^{\rm est;LQ}}\approx 0.8$. The data is $X_k=\{1,1,1,0,0,0,0,1,0,0,0,0,1,1,1\}$ and $X_k=\{1,1,0,0,0,1,1,1,1,0,0,0,1,1,1\}$, and for (a) and (b), respectively, and for the times $t_k=1.76(k-1)/(N_p-1)$ ms. }}
\label{figLQrep}
\end{figure}

\begin{figure}
\centering
\includegraphics[width=1\linewidth,angle=-0]{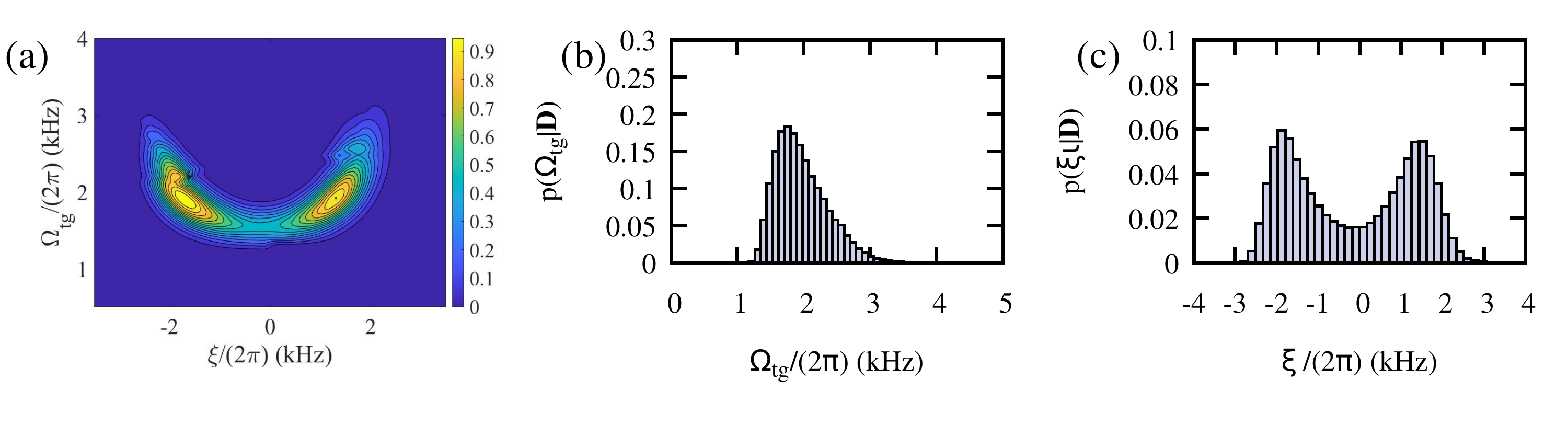}
\caption{\small{ (a) Posterior for the two unknown parameters $\xi$ and $\Omega_{\rm tg}$ when $\Omega_{\rm tg}=2\pi\times 2$ kHz and $\xi=-2\pi\times 1.5$ kHz with $B_z=1$ mT, for a realization with $N_m=4$ and $N_p=20$. Panels (b) and (c) shown the marginals for each of these parameters, which clearly reveals a bi-modal distribution for $\xi$. See main text for further details. In this case we find $\Omega_{\rm tg}^{\rm est}=2\pi\times 1.97(39)$ kHz. }}
\label{figcaseIIbi}
\end{figure}

We find that least-squares fits yield, in average, a less accurate estimator $\Omega_{\rm tg}^{\rm est}$ than its Bayesian counterpart. For that we simulate $100$ realizations where we arbitrarily chose $\Omega_{\rm tg}=2\pi\times 1.6$ kHz. For $N_m=1$ (single shot measurements) and $N_{\rm p}=15$  we find that, in average, $\overline{\delta \Omega_{\rm tg}^{\rm est;Bayes}/\delta \Omega_{\rm tg}^{\rm est;LQ}}\approx 0.8$ in the region $\Omega_{\rm tg}\in 2\pi \times \{ 1,2\}$ kHz, where LQ stands for least-squares fit. Similar results are also observed, in average, for other cases considered in the main text, namely, $N_m=N_p=15$ and $N_m=20$ with $N_p=15$ and for different values of $\Omega_{\rm tg}$. Yet, it is worth remarking that this holds in average, so it is still possible that the least-squares fit gives a more precise estimator than the Bayesian analysis for a particular realization. In order to remark this point, we plot in Fig.~\ref{figLQrep} the inverse of the sum of the residuals $\epsilon_i$, i.e. $S^{-1}\equiv (\sum_{i=1}^{N_p}\epsilon_i^2)^{-1}$, together with the posterior distribution obtain for two different realizations with $N_m=1$ and $N_p=15$. In Fig.~\ref{figLQrep}(a), the data {\bf D} is such that $S$ exhibits a global minimum at $\Omega_{\rm tg}^{\rm est;LQ}\approx 2\pi\times 0.75$ kHz, while the posterior distribution $p(\Omega_{\rm tg}|{\bf D})$ clearly reveals a peak around the true value $\Omega_{\rm tg}=2\pi\times 1.6$ kHz. In Fig.~\ref{figLQrep}(b) we show $S^{-1}$ and $p(\Omega_{\rm tg}|{\bf D})$ for a particular realization in which the least-square fit gives a more precise estimator  than through $p(\Omega_{\rm tg}|{\bf D})$ (see caption for further details). Finally, we remark that the Bayesian analysis provides the posterior distribution over the parameter of interest which contains more information than just a single estimator $\Omega_{\rm tg}^{\rm est}$ that a FFT or least-squares fit output.

 For the case II however there is no simple relation between $P_D(t)$ and $\Omega_{\rm tg}$ and $\xi$. This challenges the identification of $\Omega_{\rm tg}$ through a FFT analysis. Indeed, as shown in Fig.~\ref{figFFT}(b), the FFT spectrum reveals relevant contributions at different frequencies in a broad range of frequencies (from $5$ to $20$ kHz). Recall that $\Omega_{\rm tg}=2\pi\times 12$ kHz for this data ${\bf D}$.  Moreover, this FFT analysis cannot identify potential detunings $\xi$ w.r.t. the resonant condition. Compare this analysis with the accurate results presented in the Fig. 3 of the main text using Bayesian inference.

 One may still rely on least-squares methods aiming to determine the unknown parameters, although now the data must be fitted to the numerically-computed expression $P_D(t)=\left|\langle D|U(t,0)|D\rangle\right|^2$, where $U(t,0)=\mathcal{T}e^{-i\int_0^tds H_{\rm r}(s)}$ denotes the time evolution propagator of the time-dependent Hamiltonian $H_{\rm r}$, given in Eq. (A1) of the main text. Recall that the Hamiltonian $H_{\rm r}$ depends on these unknown parameters $\{\Omega_{\rm tg},\xi \}$. Such non-linear fit can be performed using the subroutine \texttt{lsqcurvefit} of MATLAB. In general, the fit is not capable to modify the required starting values, as it happens when choosing  $\Omega_{\rm tg}=2\pi \times 8$ kHz and $\xi=2\pi\times 0.1$ kHz as initial values (rather close to the ideal frequencies $12$ and $0.1$ kHz, respectively). From Bayesian inference, we know that these observations are more compatible with a negative detuning, so we choose a different initial pair of values, $\Omega_{\rm tg}=2\pi \times 8$ kHz and $\xi=2\pi\times -0.1$ kHz, but the fit is again incapable of finding the good solution found with our method (cf. main text), and it leads to $\Omega_{\rm tg}=2\pi \times 7.627$ kHz and $\xi=2\pi\times -0.0998$ kHz, far from the Rabi frequency of $12$ kHz. Moreover,  even when starting close to the solution, slightly different initial values lead to different results, e.g. $\Omega_{\rm tg}=2\pi\times 12.73$ kHz and $\xi=2\pi\times -0.0914$ kHz when starting from $\Omega_{\rm tg}=2\pi\times 15$ kHz, $\xi=2\pi\times 0.1$, while one obtains $\Omega_{\rm tg}=2\pi\times 11.25$ kHz and $\xi=2\pi\times 0.94$ kHz when starting from $\Omega_{\rm tg}=2\pi\times 12$ kHz and $\xi=2\pi\times 1$. This holds for other realizations, while  our Bayesian inference provides good estimates.  This further demonstrates the advantage of Bayesian inference.  In particular, for other $5$ different realizations with $N_m=4$ and $N_{p}=20$, we obtain $\Omega_{\rm tg}^{\rm est}=2\pi\times 11.79(18)$ kHz, $\xi^{\rm est}=2\pi\times 0.097(45)$ kHz, $\Omega_{\rm tg}^{\rm est}=2\pi\times 11.95(16)$ kHz, $\xi^{\rm est}=2\pi\times 0.152(35)$ kHz, $\Omega_{\rm tg}^{\rm est}=2\pi\times 11.48(18)$ kHz, $\xi^{\rm est}=2\pi\times 0.172(52)$ kHz, $\Omega_{\rm tg}^{\rm est}=2\pi\times 11.87(17)$ kHz, $\xi^{\rm est}=2\pi\times 0.118(46)$ kHz and $\Omega_{\rm tg}^{\rm est}=2\pi\times 12.09(14)$ kHz, $\xi^{\rm est}=2\pi\times 0.178(32)$ kHz, for each of the $5$ different realizations.

 Finally, and in order to emphasize the suitability of Bayesian techniques over other methods (FFT and least-squares fits) we consider a different case study, namely, $B_z=1$ mT where the amplitude is $\Omega_{\rm tg}=2\pi \times 2$ kHz and a large detuning $\xi=-2\pi\times 1.5$ kHz. Again, we take $N_m=4$ measurements for each of the $N_p=20$ points at time $t_k=0.25(k-1)/(N_p-1)$ ms. For this case, the prior for is a Gaussian centered at zero and $\sigma_\xi=2\pi\times 2$ kHz. A typical posterior obtained through the Bayesian inference is shown in Fig.~\ref{figcaseIIbi}. The marginal $p(\xi|{\bf D})$ exhibits a bi-modal structure (cf. Fig.~\ref{figcaseIIbi}(c)), which simply cannot be tackled by standard methods (FFT or least-squares fits). Even scanning the sum of the residuals $S$ for each pair of values $\Omega_{\rm tg}$ and $\xi$, the minimum will always give a single value for each of them, regardless of the distribution. In particular, for a realization of this case we find $\Omega_{\rm tg}^{\rm est;LQ}\approx 2\pi\times 1.40$ kHz, while the Bayesian inference leads to $\Omega_{\rm tg}^{\rm est;Bayes}=2\pi\times 1.81(46)$ kHz.

\end{document}